\documentclass[aps,prb,reprint,superscriptaddress,longbibliography,floatfix]{revtex4-2}

\newcommand{\boldS}{\mathbf{S}}

\usepackage{amsmath}
\usepackage{amssymb}
\usepackage{bm}
\usepackage{graphicx}
\usepackage{tikz}
\usepackage{algorithm}
\usepackage{algpseudocode}
\usetikzlibrary{arrows.meta,calc}

\begin{document}

\title{Error-Adaptive Quasi-local Decoding of the Toric Code}

\author{Hossein Dehghani}

\altaffiliation{Current address: QuEra Computing Inc., Boston, MA 02135, USA}
\affiliation{Department of Physics, University of Maryland, College Park, Maryland 20742, USA}
\affiliation{Joint Quantum Institute, NIST/University of Maryland, College Park, Maryland 20742, USA}
\affiliation{Joint Center for Quantum Information and Computer Science, NIST/University of Maryland, College Park, Maryland 20742, USA}

\author{Sarang Gopalakrishnan}
\affiliation{Department of Electrical and Computer Engineering, Princeton University, Princeton, New Jersey 08544, USA}

\author{Michael Gullans}
\affiliation{Joint Center for Quantum Information and Computer Science, NIST/University of Maryland, College Park, Maryland 20742, USA}
\affiliation{National Institute of Standards and Technology, Gaithersburg, Maryland 20899, USA}

\begin{abstract}
Topological quantum memories need decoders that are both reliable and scalable, but these goals compete: globally informed decoders are accurate near threshold yet expensive, while strictly local rules are fast but can miss long-range structure. Motivated by recent recoverability and mixed-state viewpoints, we make this tradeoff operational for the dephased toric code through a decoder-level local recoverability diagnostic. We compare global MWPM corrections to quasi-local corrections inside a target region and define a matching ratio $R_{\mathrm{match}}$, with mismatch $\epsilon_{\mathrm{match}}=1-R_{\mathrm{match}}$. Across geometry families, $\epsilon_{\mathrm{match}}$ shows strong buffer-controlled suppression and is well organized by a two-geometry scaling form. For scaled families, especially $a=b=d/8$, $R_{\mathrm{match}}$ exhibits a clear crossing and finite-size collapse near $p\!\sim\!0.09$, consistent with a growing recoverability scale near the decoding transition. We use this scaling to formulate an adaptive buffer-selection rule and to identify a distance-scaled initialization for a composite quasi-local RG decoder on the full torus. In the tuned family $a_0=b_0=d/8$, the resulting logical-failure curves show an apparent finite-size crossing at $p\simeq0.09$--$0.10$, below the conventional MWPM threshold scale. We focus on dephasing noise with perfect syndrome measurements to cleanly isolate the underlying behavior.
\end{abstract}

\maketitle

\section{Introduction}

Topological quantum error-correcting codes are a leading route to fault-tolerant quantum computation, because logical information is stored nonlocally while error syndromes are extracted through local stabilizer measurements \cite{kitaev2003fault,dennis2002topological}. In this framework, decoding is the computational step that maps syndrome data to physical corrections and therefore directly sets the practical performance-cost tradeoff of a code family. A central challenge is the tension between locality and reliability: global decoders can capture long-range correlations near threshold but are computationally costly, while local rules can be much faster but may fail precisely where long-range structure becomes important. For the dephased toric code \cite{kitaev2003fault}, global minimum-weight perfect matching (MWPM) remains the standard reference \cite{dennis2002topological,wang2003confinement}, with threshold behavior connected to the random-bond Ising transition \cite{dennis2002topological,merz2002two,harrington2004analysis}.

Recent works on mixed-state topological phases and noisy quantum memories recast this decoder tradeoff in terms of recoverability for mixed states \cite{cong2024enhancing,sang2024mixed,sang2024stability,demarti2024decoding}. Under noise and syndrome conditioning, the relevant object is a density matrix rather than a pure-code state, and the core question becomes recoverability: can logical information be reconstructed from local data, or is system-scale information fundamentally required? In this language, decodability is associated with a phase where recovery remains quasi-local up to a finite length scale, while approaching the transition drives that scale upward.

This viewpoint is closely tied to finite Markov length, local reversibility, and channel-circuit formulations of recovery \cite{sang2024stability,sang2025reversibility,lessa2025higherform,ma2025circuit}. It is also naturally connected to Petz-map and approximate-recovery theory \cite{petz1986sufficient,barnum2002reversing,fawzi2015quantum,junge2018universal,beigi2016decoding}, where one asks when local reduced information is sufficient to reconstruct the global logical content. From a decoder-design perspective, this suggests that a practically useful quantity is not only a threshold number, but a geometry-dependent recoverability scale that indicates how much neighborhood information is needed for local decisions to match global ones.

The objective of this paper is to make that recoverability-scale idea operational for a concrete decoder family. We ask: for a target region $A$, how much of the global MWPM decision inside $A$ is already determined by syndrome data in a finite neighborhood $A\cup B$? We quantify this by a matching ratio between local and global corrections \cite{sang2024mixed}, and use it as a decoder-level proxy for local recoverability.

Our main results are threefold. First, the local-global mismatch $\epsilon_{\mathrm{match}}=1-R_{\mathrm{match}}$ follows a simple scaling structure in $(a,b)$ with strong buffer-controlled suppression. Second, for distance-scaled families (notably $a=b=d/8$ in our data), the matching ratio exhibits a clear crossing and finite-size collapse near $p\sim 0.09$, consistent with a growing recoverability scale near the transition. Third, this scaling yields an adaptive buffer rule and motivates the distance-scaled initialization of a composite quasi-local RG decoder on the full torus. For the families studied here, that composite decoder displays threshold-like logical-failure behavior, with a finite-size crossing below the conventional MWPM threshold scale.

We intentionally keep the scope narrow. We focus on dephasing with perfect measurements so the matching-ratio mechanism can be isolated cleanly. Extensions to repeated-syndrome noisy-measurement decoding \cite{hauser2026information,caune2024realtime}, and to alternative fast local solver families \cite{delfosse2021almost,chan2023actis,wolanski2025ambiguity}, are left for separate treatment.

The remainder of the paper is organized as follows. In Sec.~II we define the subregion decoder and the matching ratio. Section~III introduces the scaling ansatz for the local-global mismatch, and Sec.~IV shows the transition-like behavior of the matching ratio for scaled families. In Sec.~V we extract the adaptive buffer rule, in Sec.~VI we construct the composite quasi-local RG decoder, in Sec.~VII we present its whole-system logical-failure results, in Sec.~VIII we analyze complexity and fast-decoding implications, and in Sec.~IX we discuss broader implications and extensions.

\section{Subregion decoder and matching ratio}

We begin by fixing the toric-code geometry, the target and buffer regions, and the operational definition of the matching ratio.

The toric code and its boundary (surface-code) variant are canonical topological stabilizer-code constructions \cite{kitaev2003fault,bravyi1998surface}. The toric code places physical qubits on edges of a square lattice with periodic boundary conditions. Using the standard convention, the stabilizers are
\begin{equation}
A_v=\prod_{e\ni v} X_e,
\qquad
B_p=\prod_{e\in\partial p} Z_e,
\label{eq:toric_stabilizers}
\end{equation}
where $A_v$ is the star operator at vertex $v$ and $B_p$ is the plaquette operator on face $p$. The code space is the simultaneous $+1$ eigenspace of all $A_v$ and $B_p$.

Logical operators are noncontractible loops around the two cycles of the torus. A convenient representative set is
\begin{equation}
\bar{Z}_{x/y}=\prod_{e\in\gamma_{x/y}} Z_e,
\qquad
\bar{X}_{x/y}=\prod_{e\in\tilde{\gamma}_{x/y}} X_e,
\end{equation}
where $\gamma_{x/y}$ and $\tilde{\gamma}_{x/y}$ are primal and dual noncontractible cycles.

In the dephasing setting, physical errors are $Z$-type, so the relevant syndrome is generated by violated $X$-type checks (star defects), while $Z$ plaquette checks remain unaffected by the noise channel. Decoding therefore reduces to pairing these syndrome defects on the vertex lattice; throughout this manuscript, the global reference decoder is MWPM on the full torus.

For this dephasing code-capacity model (independent $Z$ errors, perfect syndrome measurements), the natural optimal benchmark is maximum-likelihood decoding (MLD). Here MLD means choosing the most likely logical class of syndrome-compatible error chains, rather than reconstructing one microscopic error chain exactly. For a fixed syndrome, many $Z$-error chains produce the same defect pattern, and two such chains can differ either by local contractible loops or by noncontractible loops that change the encoded logical action. Successful decoding therefore depends on whether the final error-plus-correction is topologically trivial on the torus, not on whether the decoder reproduces the physical error chain edge by edge. The MLD groups all compatible chains according to their logical action, sums their probabilities within each class, and chooses the most probable class.

Exact MLD is, however, computationally intractable in general: optimal decoding of stabilizer codes is \#P-complete \cite{iyer2015hardness}, and identifying a most likely compatible error is NP-hard \cite{hsieh2011nphard}. Practical decoders such as MWPM, and the composite decoder studied here, are therefore suboptimal surrogates for an intractable optimum.

In the toric-code setting this class-summed decoding problem maps to the phase transition of the two-dimensional random-bond Ising model (RBIM) on the Nishimori line. This mapping was established in the Dennis--Kitaev--Landahl--Preskill work \cite{dennis2002topological} and analyzed further in Refs.~\cite{wang2003confinement,honecker2001universality}. It gives a natural reference scale $p_c^{\mathrm{ML}}\approx 0.109$ (about $10.9\%$), while practical MWPM thresholds for the same setting are typically slightly lower, around $p_c^{\mathrm{MWPM}}\approx 0.103$ \cite{dennis2002topological,wang2003confinement}. We use these values as baseline context for the composite-decoder results reported below.

\begin{figure}[tbp]
    \centering
\begin{tikzpicture}[x=0.48cm,y=0.48cm,>=Stealth]
    \def\L{12}      
    \def\Ax{4}      
    \def\Ay{4}
    \def\As{4}      
    \def\bw{2}      

    \pgfmathsetmacro{\Axl}{\Ax}
    \pgfmathsetmacro{\Ayl}{\Ay}
    \pgfmathsetmacro{\Axu}{\Ax+\As}
    \pgfmathsetmacro{\Ayu}{\Ay+\As}
    \pgfmathsetmacro{\Bxl}{\Ax-\bw}
    \pgfmathsetmacro{\Byl}{\Ay-\bw}
    \pgfmathsetmacro{\Bxu}{\Ax+\As+\bw}
    \pgfmathsetmacro{\Byu}{\Ay+\As+\bw}

    \path[fill=red!70!black, fill opacity=0.07, even odd rule]
        (\Bxl,\Byl) rectangle (\Bxu,\Byu)
        (\Axl,\Ayl) rectangle (\Axu,\Ayu);
    \path[fill=blue!70!black, fill opacity=0.07]
        (\Axl,\Ayl) rectangle (\Axu,\Ayu);

    \foreach \i in {0,...,\L} {
        \draw[black!15,line width=0.25pt] (\i,0) -- (\i,\L);
        \draw[black!15,line width=0.25pt] (0,\i) -- (\L,\i);
    }

    \foreach \i in {0,...,11} {
        \foreach \j in {0,...,\L} {
            \fill[black] ({\i+0.5},\j) circle (0.055);
            \fill[black] (\j,{\i+0.5}) circle (0.055);
        }
    }

    \draw[black, line width=0.9pt] (0,0) rectangle (\L,\L);
    \draw[red!70!black, line width=1.0pt] (\Bxl,\Byl) rectangle (\Bxu,\Byu);
    \draw[blue!70!black, line width=1.0pt] (\Axl,\Ayl) rectangle (\Axu,\Ayu);

    \node[blue!70!black] at ({(\Axl+\Axu)/2},{(\Ayl+\Ayu)/2}) {$A$};
    \node[red!70!black] at ({(\Bxl+\Bxu)/2},{\Byu-0.6}) {$B$};

    \draw[<->, line width=0.8pt]
        (0,{\L+0.7}) -- (\L,{\L+0.7})
        node[midway, fill=white, inner sep=1pt] {$2d$};
    \draw[<->, line width=0.8pt]
        ({\Axl-0.7},\Ayl) -- ({\Axl-0.7},\Ayu)
        node[midway, fill=white, inner sep=1pt] {$a$};
    \draw[<->, line width=0.8pt]
        ({\Bxl+0.8},\Ayu) -- ({\Bxl+0.8},\Byu)
        node[midway, fill=white, inner sep=1pt] {$b$};
\end{tikzpicture}
    \caption{Geometry of the local decoder region for the toric code (qubits live on edges, shown as filled circles). The local decoder acts on a target square $A$ of linear size $a$ (blue) together with a surrounding buffer region $B$ of width $b$ (red). The full system is a distance-$d$ toric code with periodic boundary conditions (not shown), drawn on the underlying $2d\times 2d$ grid; one $X$-stabilizer (vertex) lattice spacing corresponds to two grid units, and $a$ and $b$ are quoted in vertex-lattice spacings throughout the paper.}
    \label{fig:geometry_rewrite}
\end{figure}
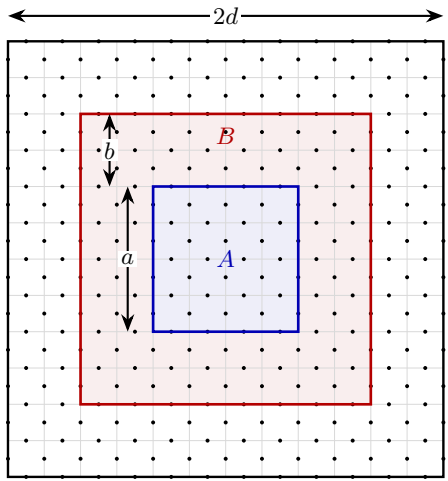

To define a local decoder, we consider a square target region $A$ with linear size $a$, together with an annulus buffer region $B$ with width $b$ around it (Fig.~\ref{fig:geometry_rewrite}). The local decoder has access to the syndrome in $A\cup B$, but the correction that is kept is restricted to the target region $A$. Throughout the paper, $a$ and $b$ are measured in units of the $X$-stabilizer (vertex) lattice spacing: a distance-$d$ toric code corresponds to a $d\times d$ periodic vertex lattice, while the geometry schematic is drawn on the underlying $2d\times 2d$ grid containing vertices, plaquette centers, and edge qubits. In that drawing, one vertex-lattice spacing corresponds to two grid units. Every later scaling plot and every RG step in the composite decoder refers back to this distinction: corrections are committed only in the target region, while the buffer supplies auxiliary syndrome information.

Given two decoders, we define the matching ratio of their outputs as the fraction of samples for which the two decoders give the same correction inside the target region. More explicitly, let $\mathcal{D}_{\mathrm{glob}}$ be the global MWPM decoder and let $\mathcal{D}_{A,B}$ be the local decoder acting on $A\cup B$. For a syndrome sample $\boldS_i$, let $P_{i,A}^{\mathrm{glob}}$ be the restriction of the global Pauli correction to $A$, and let $P_{i,A}^{A,B}$ be the Pauli correction produced by the local decoder in $A$. Then
\begin{equation}
R_{\mathrm{match}}(\mathcal{D}_{\mathrm{glob}},\mathcal{D}_{A,B})
=
\frac{1}{N}\sum_{i=1}^{N}\delta\!\left(P_{i,A}^{\mathrm{glob}},\;P_{i,A}^{A,B}\right),
\label{eq:matching_ratio}
\end{equation}
where $\delta(x,y)=1$ if $x=y$ and $0$ otherwise.

In this manuscript, equality in Eq.~(\ref{eq:matching_ratio}) is literal Pauli-string equality on $A$ for the fixed decoder implementations and deterministic tie-breaking used in the simulations. We do not quotient by local stabilizers inside $A$, nor do we reduce the comparison to a class-only notion of equivalence. Such a stabilizer-quotiented or class-only statistic would be the more natural object for probing the locality structure of MLD itself, since MLD is defined at the level of logical equivalence classes rather than specific correction representatives. In practice, however, the decoders studied here return concrete Pauli representatives, so we use literal representative agreement as the operational statistic relevant for decoder design.

This quantity is useful for two reasons. First, it directly measures how much of the global pairing decision is already fixed by local syndrome information. Second, it is much cheaper to scan over many choices of $a$ and $b$ using the matching ratio than to build a full whole-system decoder every time. So in practice the matching-ratio analysis is the first step in deciding how to build the composite decoder.

A statistic of this type was introduced in Ref.~\cite{sang2024mixed}, where the probability that a truncated MWPM solve on a block plus surrounding buffer reproduces the global pairing on the block (with block, buffer, and system size scaled together) is used to establish the toric-code mixed-state phase. Our $R_{\mathrm{match}}$ is its decoder-oriented counterpart: the target and buffer sizes are varied independently, agreement is literal equality of the committed correction inside $A$, and the statistic is used to select decoder geometry rather than to construct a local channel for phase equivalence.

The matching ratio has the same broad flavor as more information-theoretic recoverability diagnostics, especially the conditional mutual information (CMI) used to define a finite Markov length \cite{sang2024stability}; see also Refs.~\cite{sang2025reversibility,ma2025circuit}. To make that comparison precise, let $C$ denote the exterior complement of $A\cup B$ on the torus, so that $A$ is the target region, $B$ is the surrounding buffer, and $C$ is the rest of the system beyond the buffer. The CMI is
\begin{equation}
I(A\!:\!C|B)=S(AB)+S(BC)-S(B)-S(ABC),
\end{equation}
with $S(X)=-\mathrm{Tr}\,\rho_X\log\rho_X$ the von Neumann entropy of region $X$.

The point of invoking CMI here is not the entropy formula by itself, but its conditional-information meaning. Small $I(A\!:\!C|B)$ means that once the information in the buffer region $B$ is known, the exterior region $C$ contains little additional information about $A$; equivalently, $B$ approximately screens $A$ from $C$, so $A$-$B$-$C$ behaves like an approximate Markov chain. This is conceptually close to what $R_{\mathrm{match}}$ asks operationally. There too one conditions on the local data in $A\cup B$ and asks whether access to the outside of the buffer still changes the decision inside $A$. In our decoder language, a small mismatch $\epsilon_{\mathrm{match}}=1-R_{\mathrm{match}}$ means that the exterior region $C$ rarely changes the correction selected in $A$ once the decoder already sees $A\cup B$. Thus the analogy is about conditional screening by the buffer, not about identifying entropies with decoder outputs. The two quantities are therefore related in interpretation but not identical: $I(A\!:\!C|B)$ is a state property with an information-theoretic recovery meaning, whereas $R_{\mathrm{match}}$ and $\epsilon_{\mathrm{match}}$ are explicitly decoder-dependent agreement statistics.

\section{Scaling ansatz for the matching ratio}

Having defined the decoder observable, we next ask how it depends on geometry and error rate.

For decoder design, it is convenient to work with the mismatch variable
\begin{equation}
\epsilon_{\mathrm{match}}(p;a,b)=1-R_{\mathrm{match}}(p;a,b).
\label{eq:epsilon_match}
\end{equation}
This local-global mismatch is easier to organize across geometry families than $R_{\mathrm{match}}$ itself.

A form that works well for the present data is
\begin{equation}
\epsilon_{\mathrm{match}}(p;a,b;d)
\simeq
C(p,d)\left(\frac{a}{b}\right)^{\alpha(p)} e^{-b/\xi(p)},
\label{eq:epsilon_ansatz}
\end{equation}
with $C(p,d)$ an amplitude that is independent of the patch geometry $(a,b)$ but not, in general, of the system size $d$. Eq.~(\ref{eq:epsilon_ansatz}) is used as an empirical finite-size fit, not as an exact closed-form derivation. We fit it at a single distance, which determines $\alpha(p)$ and $\xi(p)$ but leaves the $d$-dependence of the amplitude unresolved; on general scaling grounds one expects $d$ to enter through the ratio $d/\xi(p)$ and to drop out once $d\gg\xi(p)$, which is the regime of the fits reported here. The structure of Eq.~(\ref{eq:epsilon_ansatz}) should be compared carefully with the CMI scaling ansatz in Ref.~\cite{sang2024stability}. In their toric-code geometry, the inner region $A$ is kept $O(1)$ while the annular width $r$ of the surrounding regions is varied, leading near criticality to a one-parameter form $I(A\!:\!C|B)=r^{-\alpha}\Phi((p-p_c)r^{1/\nu})$ and hence $I(A\!:\!C|B)\sim r^{-\alpha}$ exactly at $p=p_c$; away from criticality the same work finds exponential decay $I(A\!:\!C|B)\sim e^{-r/\xi_\pm(p)}$ \cite{sang2024stability}. This power law in the buffer size is natural in their setting because, after coarse graining, the only growing geometric scale is the annular width itself. Our ansatz is therefore of the same flavor, but not the same object and not the same scaling form. Here the target size is itself varied and can be comparable to the buffer, so the residual geometry dependence is kept explicitly through the prefactor $(a/b)^{\alpha(p)}$ rather than being absorbed into a pure power of the buffer width alone. Eq.~(\ref{eq:epsilon_ansatz}) should therefore be read as a decoder-geometry ansatz for the mismatch, not as the CMI ansatz of Ref.~\cite{sang2024stability} transplanted verbatim.

\begin{figure*}[tbp]
    \centering
    \begin{minipage}[t]{0.485\textwidth}
        \centering
        \makebox[\linewidth][l]{\hspace{0.2em}\textbf{(a)}}\\[-0.15em]
        \includegraphics[width=0.95\linewidth]{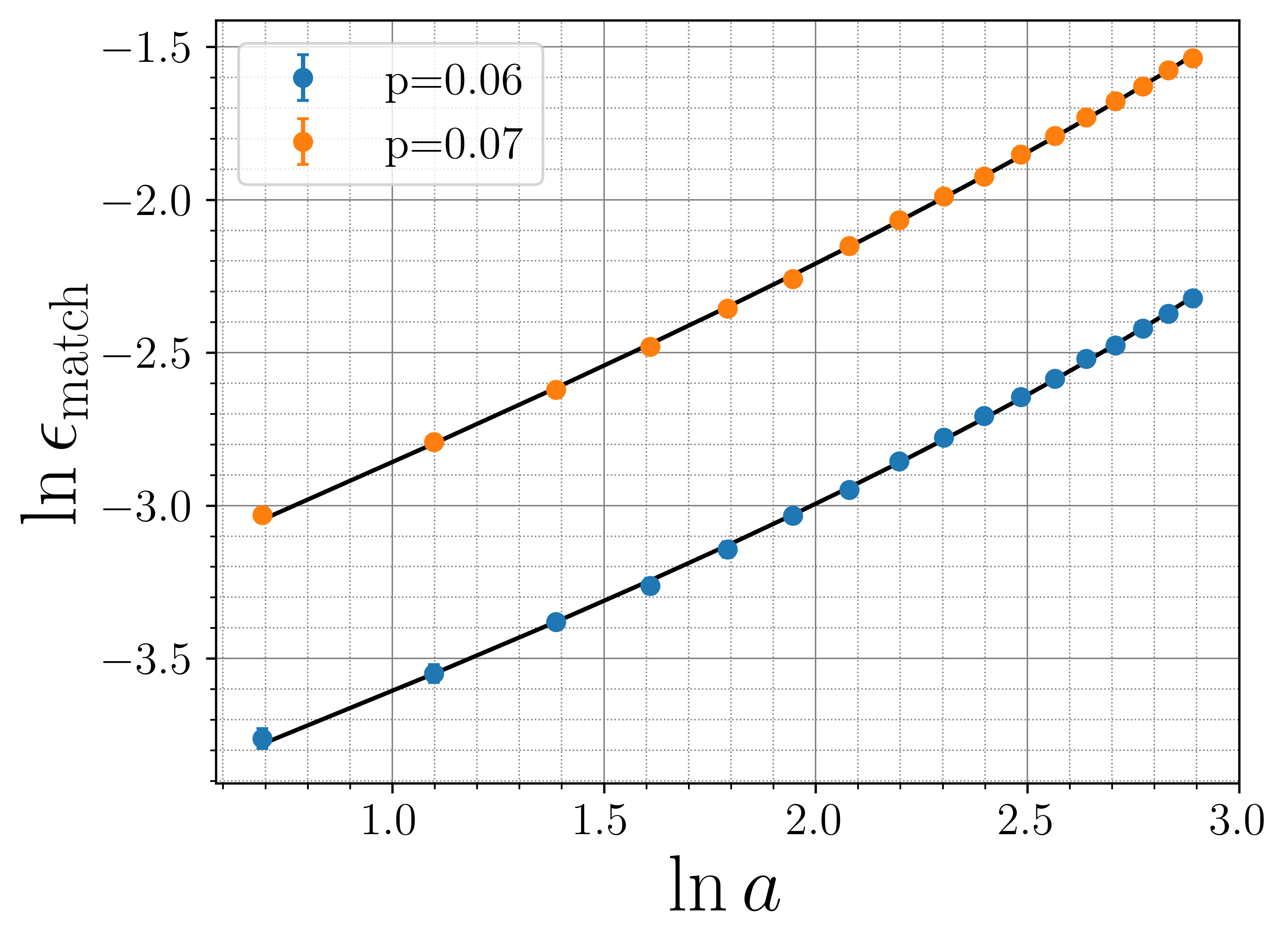}
    \end{minipage}\hfill
    \begin{minipage}[t]{0.485\textwidth}
        \centering
        \makebox[\linewidth][l]{\hspace{0.2em}\textbf{(b)}}\\[-0.15em]
        \includegraphics[width=0.95\linewidth]{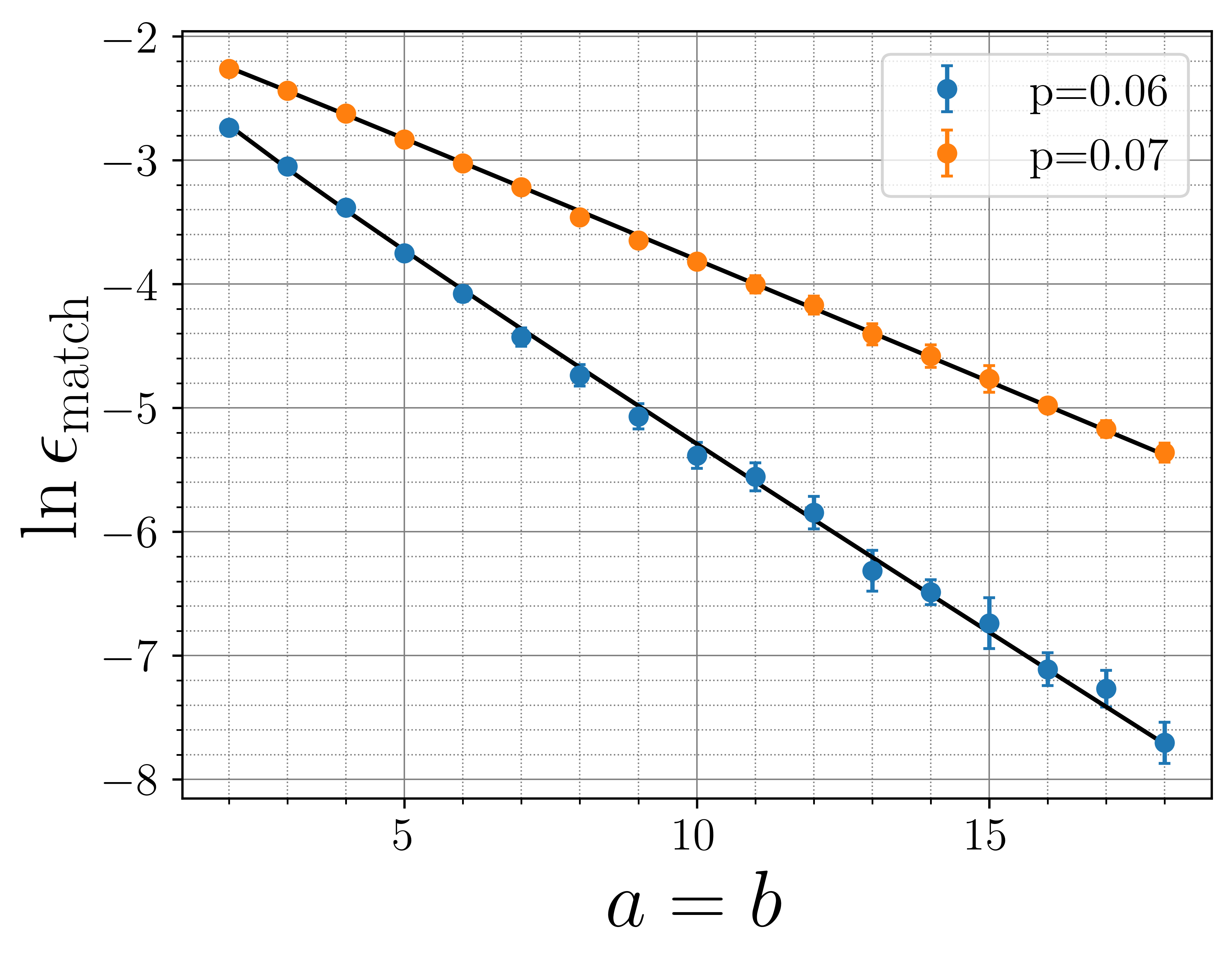}
    \end{minipage}\hfill

    \vspace{0.6em}

    \begin{minipage}[t]{0.55\textwidth}
        \centering
        \makebox[\linewidth][l]{\hspace{0.2em}\textbf{(c)}}\\[-0.15em]
        \includegraphics[width=0.95\linewidth]{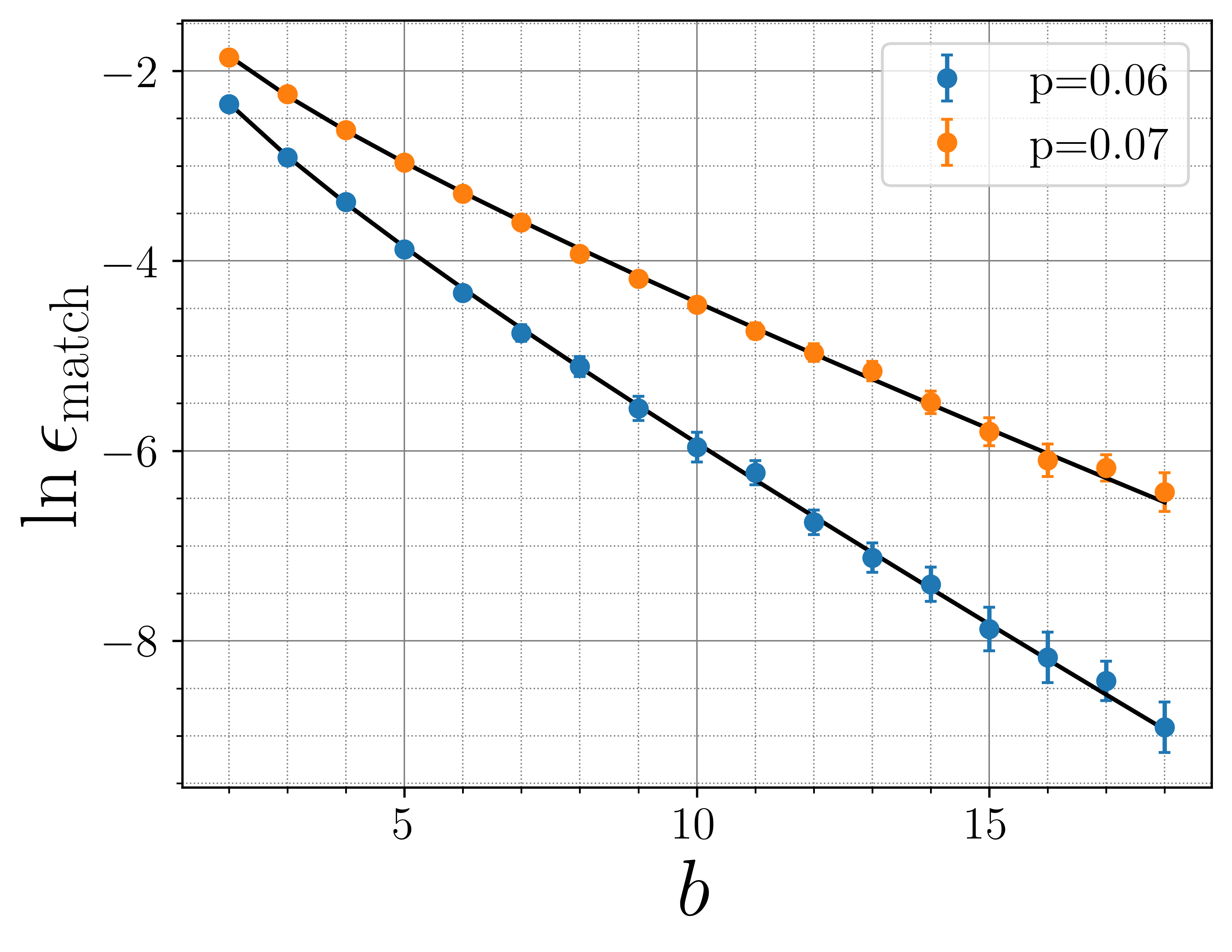}
    \end{minipage}
\caption{Numerical scaling of the local-global mismatch. (a) $\ln \epsilon_{\mathrm{match}}$ versus $\ln a$. (b) $\ln \epsilon_{\mathrm{match}}$ for the constrained family $a=b$. (c) $\ln \epsilon_{\mathrm{match}}$ versus $b$. Panel (a) isolates target-size dependence, panel (b) probes the scaled family used later in the decoder construction, and panel (c) isolates buffer-size dependence. Taken together, the three panels motivate the two-geometry ansatz in Eq.~(\ref{eq:epsilon_ansatz}) rather than a single-variable CMI-style scaling form controlled only by buffer width. All panels are at $d=64$, with the complementary length held fixed at vertex scale $4$ ($b=4$ in (a), $a=4$ in (c)); error bars are $1\sigma$ statistical. Black lines are fits to Eq.~(\ref{eq:epsilon_ansatz}), i.e. $\ln\epsilon_{\mathrm{match}}=\ln C+\alpha(p)\ln(a/b)-b/\xi(p)$, giving $\alpha\simeq0.52$ and $0.58$ at $p=0.06$ and $0.07$.}
\label{fig:epsilon_scaling}
\end{figure*}

The truncated-matching analysis of Ref.~\cite{sang2024mixed} likewise modeled the local-global disagreement as a pure exponential in the buffer width, which is natural in its tied-scale geometry, where only one length exists by construction. Once $a$ and $b$ are decoupled, a buffer-width exponential alone no longer organizes the data [Fig.~\ref{fig:epsilon_scaling}(a)]; the $(a/b)^{\alpha(p)}$ prefactor supplies the missing target-size dependence, while Eq.~(\ref{eq:epsilon_ansatz}) reduces back to that single-scale form on constrained families such as $a=b$.

The interpretation is then as follows: $\xi(p)$ is an effective recoverability length controlling buffer suppression, and $\alpha(p)$ captures the residual dependence on how large the target is relative to the buffer. In this work we use $C(p,d)$, $\alpha(p)$, and $\xi(p)$ as effective fitting parameters that organize the data and motivate the decoder design; a dedicated extraction of these parameters as physical observables is left to future work.

Representative scaling behavior is shown in Fig.~\ref{fig:epsilon_scaling}. Panel (a) isolates the target-size dependence, whose slope gives $\alpha(p)$, panel (c) isolates the suppression with buffer size, and panel (b) shows the constrained family $a=b$ that is later used as a scaled decoder family. If the mismatch were controlled by a single buffer scale, as in the CMI problem with a fixed $O(1)$ target region, panel (a) would carry little independent information once panel (c) was known. Instead, the data are organized by a genuine two-geometry dependence on both $a$ and $b$, which is why the decoder design below depends on both target size and buffer size, and why the scaled families receive separate attention. Quantitatively, fitting each panel to a power of the scanned length times the buffer exponential returns a power $+0.52$ for the target scan at fixed buffer, $-0.55$ for the buffer scan at fixed target, and $\approx0$ on the $a=b$ diagonal (values at $p=0.06$; the target-scan exponent rises to $0.58$ at $p=0.07$). This approximate $+\alpha/{-\alpha}/0$ pattern is the fingerprint of the ratio prefactor $(a/b)^{\alpha}$: the target and buffer powers are equal and opposite to within the fitting uncertainty ($+0.52$ versus $-0.55$), so the ratio form is what the data support at the present statistical confidence rather than as an exact identity. It also furnishes a near parameter-free check: the target-scan exponent $\alpha\simeq0.52$ predicts a factor $b^{-\alpha}$ in the buffer scan, and a direct fit there gives $-0.55$, consistent within uncertainty and with no adjustable parameters.

\section{Crossing and finite-size scaling of the matching ratio}

We next focus on scaled geometry families, where the matching ratio can be analyzed as a transition-like observable in its own right, connecting the subregion statistic to a growing recoverability scale near the decoding threshold.

\begin{figure*}[tbp]
    \centering
    \begin{minipage}[t]{0.495\textwidth}
        \centering
        \includegraphics[width=0.95\linewidth]{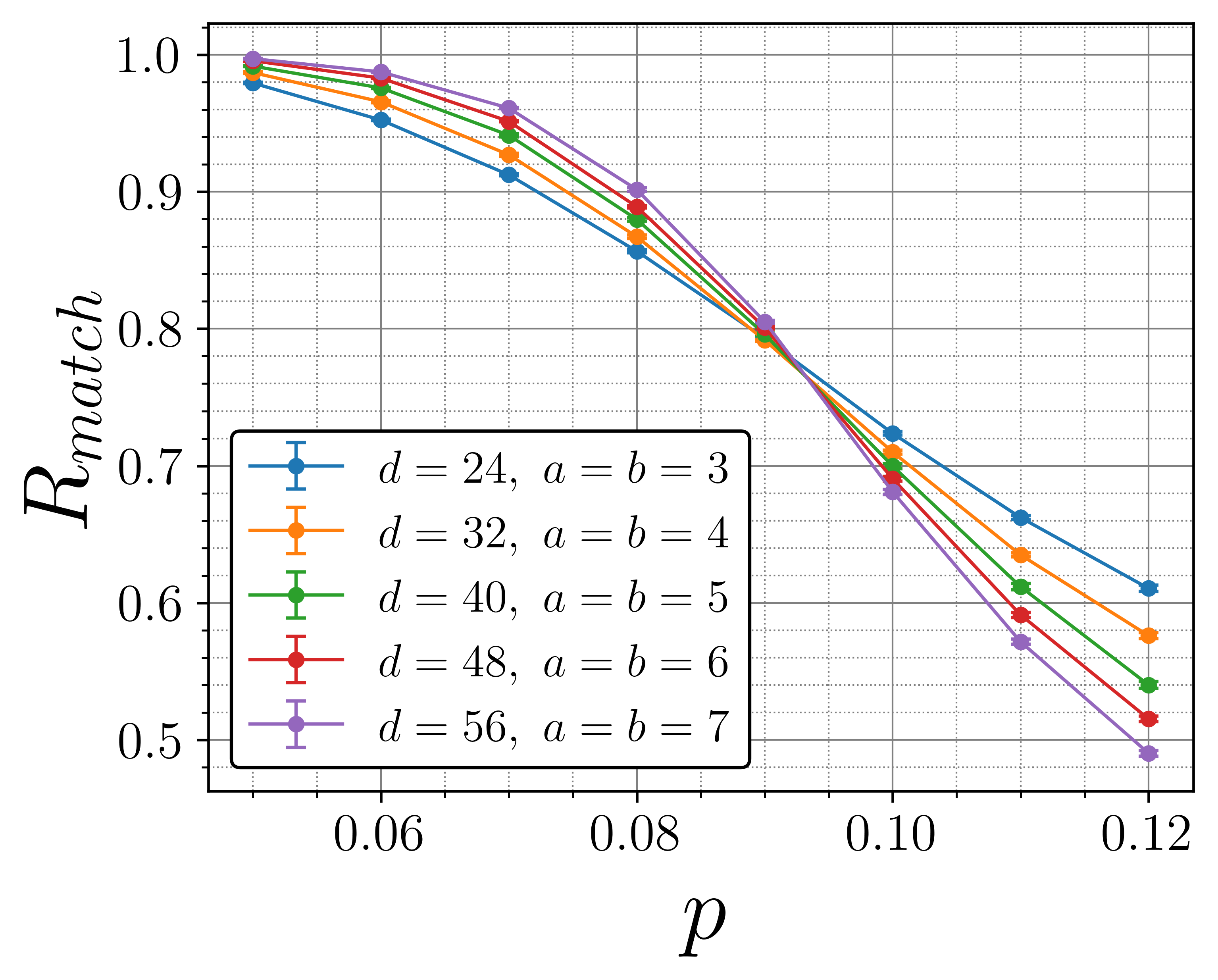}
    \end{minipage}\hfill
    \begin{minipage}[t]{0.495\textwidth}
        \centering
        \includegraphics[width=0.95\linewidth]{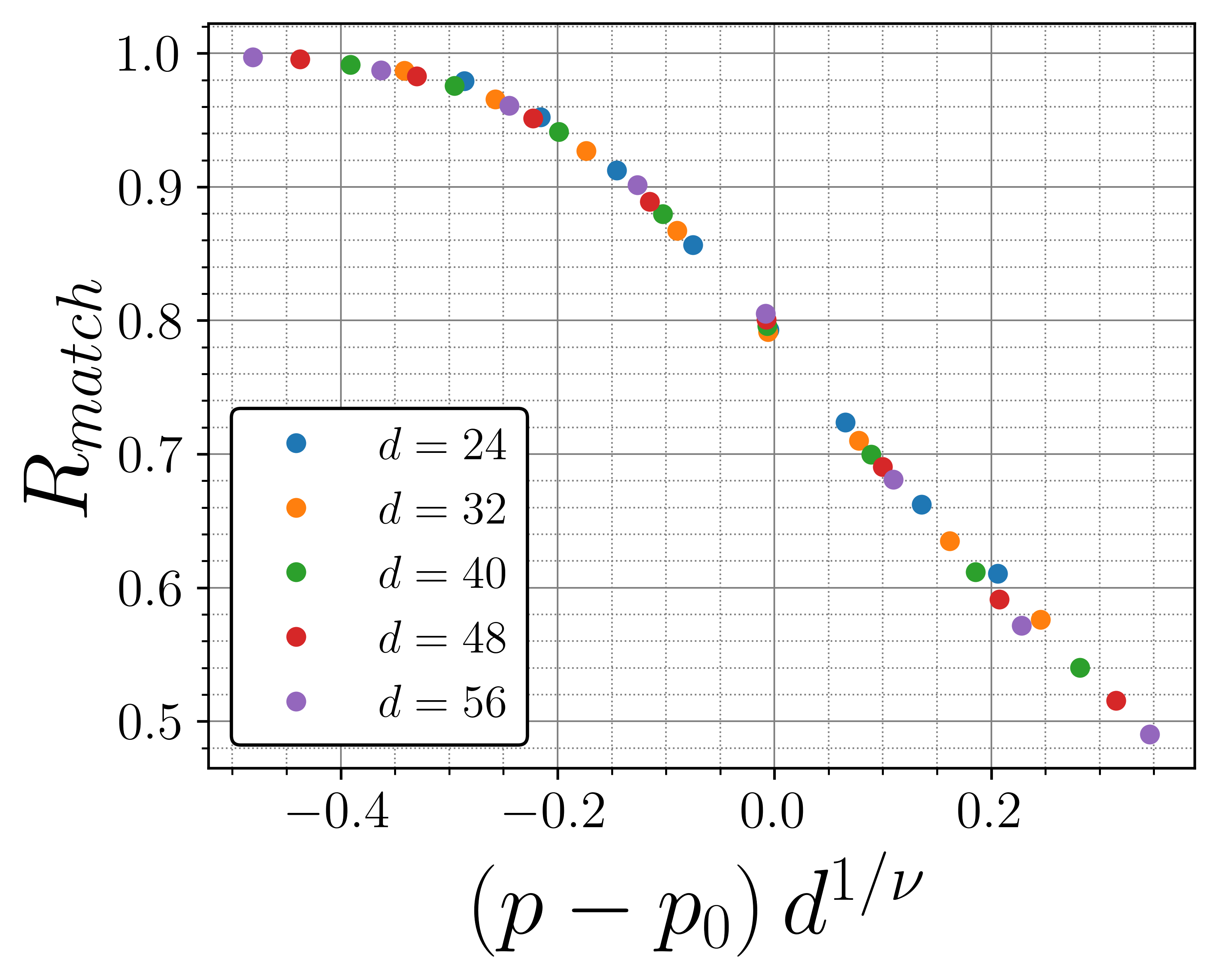}
    \end{minipage}
    \caption{Crossing and finite-size scaling of the matching ratio for the scaled family $a=b=d/8$. (a) Crossing behavior of $R_{\mathrm{match}}(p)$ across code distances. (b) Corresponding finite-size collapse around $p_0$. The crossing signals transition-like behavior of the local-global agreement once the local support is scaled with system size.}
    \label{fig:rmatch_transition}
\end{figure*}

Once $a$ and $b$ are scaled together, the matching ratio develops a transition-like structure as a function of physical error rate. In the data analyzed here, one clean family is $a=b=d/8$, shown in Fig.~\ref{fig:rmatch_transition}. The curves cross near $p_0\approx0.091$, and a simple collapse works with an effective exponent $\nu\sim1.63$ for the plotted dataset. Panel (a) provides direct evidence for a crossing in the decoder observable itself, while panel (b) shows that the same dataset can be organized by a standard one-parameter finite-size scaling variable.

First, once the local support is allowed to grow with system size, the quasi-local decoder develops a sharp transition at an error rate close to, but measurably below, the global decoding threshold. The crossing at $p_0\approx0.091$ lies below $p_c^{\mathrm{MWPM}}\approx0.103$ for the same noise model. At the present sizes we cannot cleanly separate the possible origins of this offset: residual finite-size drift of the crossing point, the fact that literal representative agreement is a stricter criterion than logical success, and the possibility that quasi-local agreement is genuinely lost before decodability itself. We therefore treat $p_0$ as a crossover scale for the matching statistic, not as the logical threshold. Second, the analysis gives a concrete way to define the local scales inside the composite decoder. The regions $A$ and $B$ are extracted from the subregion data rather than chosen by hand.

For this family, finite-size analysis is consistent with
\begin{equation}
\begin{aligned}
R_{\mathrm{match}}(p,d)&\approx
\mathcal{F}\!\left((p-p_0)d^{1/\nu}\right),\\
p_0 &\approx 0.091,
\qquad
\nu \approx 1.63.
\end{aligned}
\label{eq:rmatch_collapse}
\end{equation}
Using a fixed broad-window fitting protocol for this family, we obtain $p_0\approx0.091$ and $\nu\approx1.63$. We do not attach formal error bars to these values: the dominant uncertainty is not the statistical scatter of a single-window fit but the finite-size drift of the crossing point discussed above, which a fixed set of distances does not capture, and $\nu$ should accordingly be read as an effective collapse parameter rather than a precise exponent. The quality of the collapse is sufficient for the decoder-design use made of it below, though not for a precision determination of the exponents.

The crossing resides in the system-size dependence of the amplitude $C(p,d)$, which the fixed-distance fits of Sec.~III do not determine. On a scaled family $a=b$ the ratio prefactor is unity and Eq.~(\ref{eq:epsilon_ansatz}) reduces to $\epsilon_{\mathrm{match}}\simeq C(p,d)\,e^{-a/\xi(p)}$; were $C$ independent of $d$, the curves of the family would be strictly ordered in $d$ at every $p$ and could never cross. The crossing is thus direct evidence that $C(p,d)$ varies with system size. The reason this variation is invisible in Fig.~\ref{fig:epsilon_scaling} is one of regime rather than of geometric ratio: those scans sit well below threshold, where $\xi(p)$ is short compared with the system, the exterior of the patch lies many correlation lengths away, and the amplitude is effectively saturated. The scaled families instead cross near $p_0$, where $\xi(p)$ has grown comparable to $d$ and the amplitude is no longer saturated. The crossing is therefore a finite-size effect controlled by $d/\xi(p)$, precisely the combination organized by the collapse variable in Eq.~(\ref{eq:rmatch_collapse}); resolving the form of $C(p,d)$, expected to be a scaling function of $d/\xi(p)$, is left to future work. Empirically, the $a=b=d/2$ family in Fig.~\ref{fig:appendix_scaled_families} crosses at slightly higher $p$ than the $a=b=d/4$ family, consistent with larger local supports tracking the global decoder to higher error rates.

The current fitted value $\nu\approx1.63$ is somewhat larger than the $\nu\approx1.5$ value commonly associated with the two-dimensional RBIM/Nishimori universality class for global decoding observables \cite{merz2002two,dennis2002topological,honecker2001universality}. At the present system sizes, this difference should be read as a finite-size and observable-dependent effect rather than as evidence for a distinct universality class. A broader-size scaling study is needed for a definitive exponent estimate.

\section{Adaptive buffer selection}

The scaling analysis becomes practically useful once it is converted into an adaptive rule for choosing the buffer size. Given a target mismatch tolerance $\eta$, we define
\begin{equation}
b_c(p,\eta;\rho)=
\min\bigl\{b:\epsilon_{\mathrm{match}}(p;\rho b,b)\le \eta\bigr\},
\label{eq:bc_def}
\end{equation}
where $\rho=a/b$ is fixed within a chosen scaled family. Here $\eta$ is a user-chosen tolerance parameter, distinct from the measured mismatch $\epsilon_{\mathrm{match}}$ itself. This says how large the buffer should be at a given physical error rate if we want the local decoder to stay within a prescribed mismatch from the global one. Within the ansatz of Eq.~(\ref{eq:epsilon_ansatz}), setting $a=\rho b$ gives the closed form
\begin{equation}
b_c(p,\eta;\rho)\simeq\xi(p)\left[\ln\frac{C(p,d)}{\eta}+\alpha(p)\ln\rho\right],
\label{eq:bc_closed}
\end{equation}
rounded up to the nearest integer buffer size. Because the ratio prefactor is independent of $b$ at fixed $\rho$, this inversion is exact within the ansatz. In practice we evaluate Eq.~(\ref{eq:bc_def}) directly on the measured mismatch data rather than through the fit; Eq.~(\ref{eq:bc_closed}) makes explicit how the required buffer inherits the growth of $\xi(p)$ near the transition. We do not extract $\xi(p)$ itself in this work; the measured growth of $b_c(p,\eta)$ with $p$ plays that role operationally.

\begin{figure*}[t]
    \centering
    \begin{minipage}[t]{0.495\textwidth}
        \centering
        \includegraphics[width=0.95\linewidth]{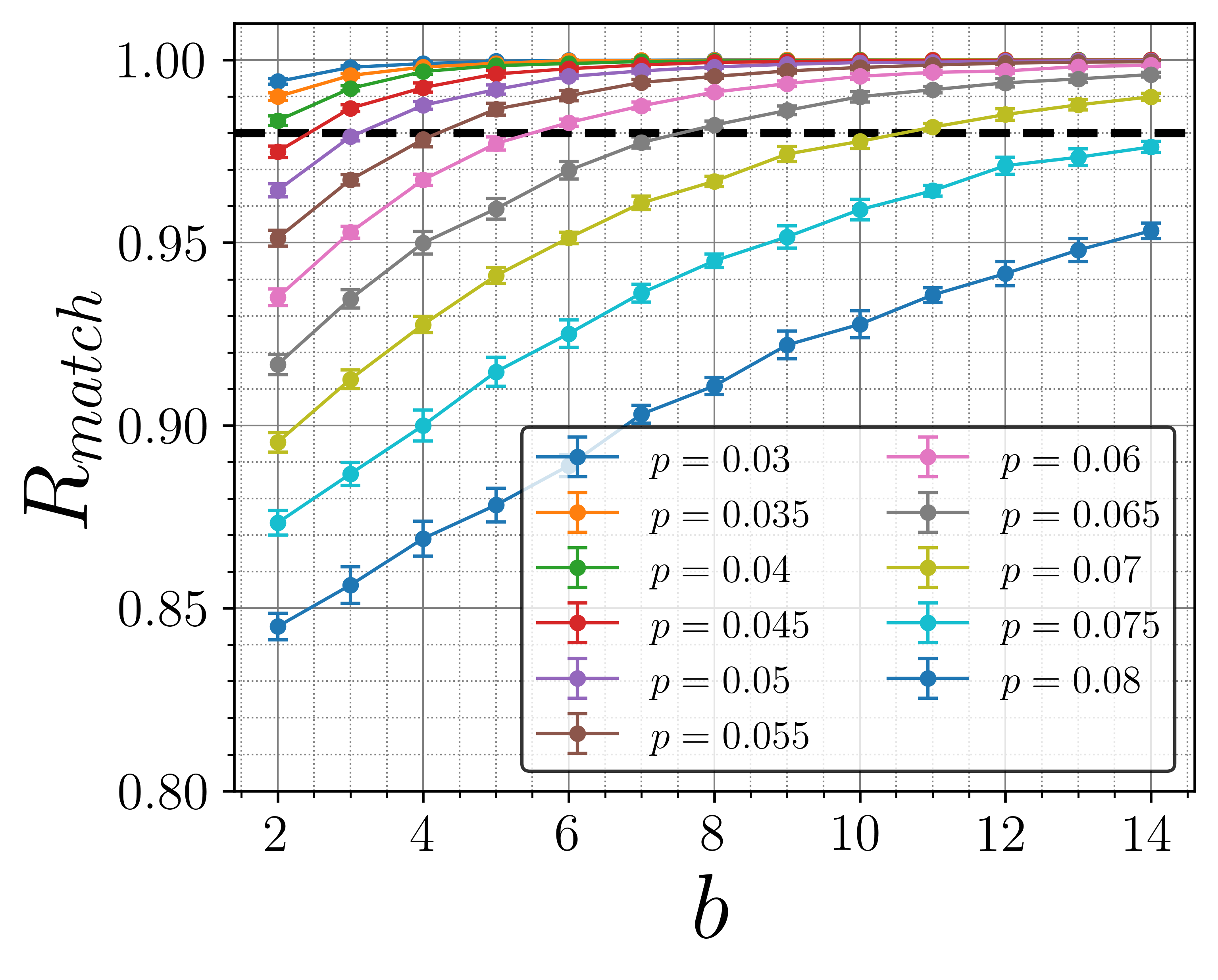}
    \end{minipage}\hfill
    \begin{minipage}[t]{0.495\textwidth}
        \centering
        \includegraphics[width=0.95\linewidth]{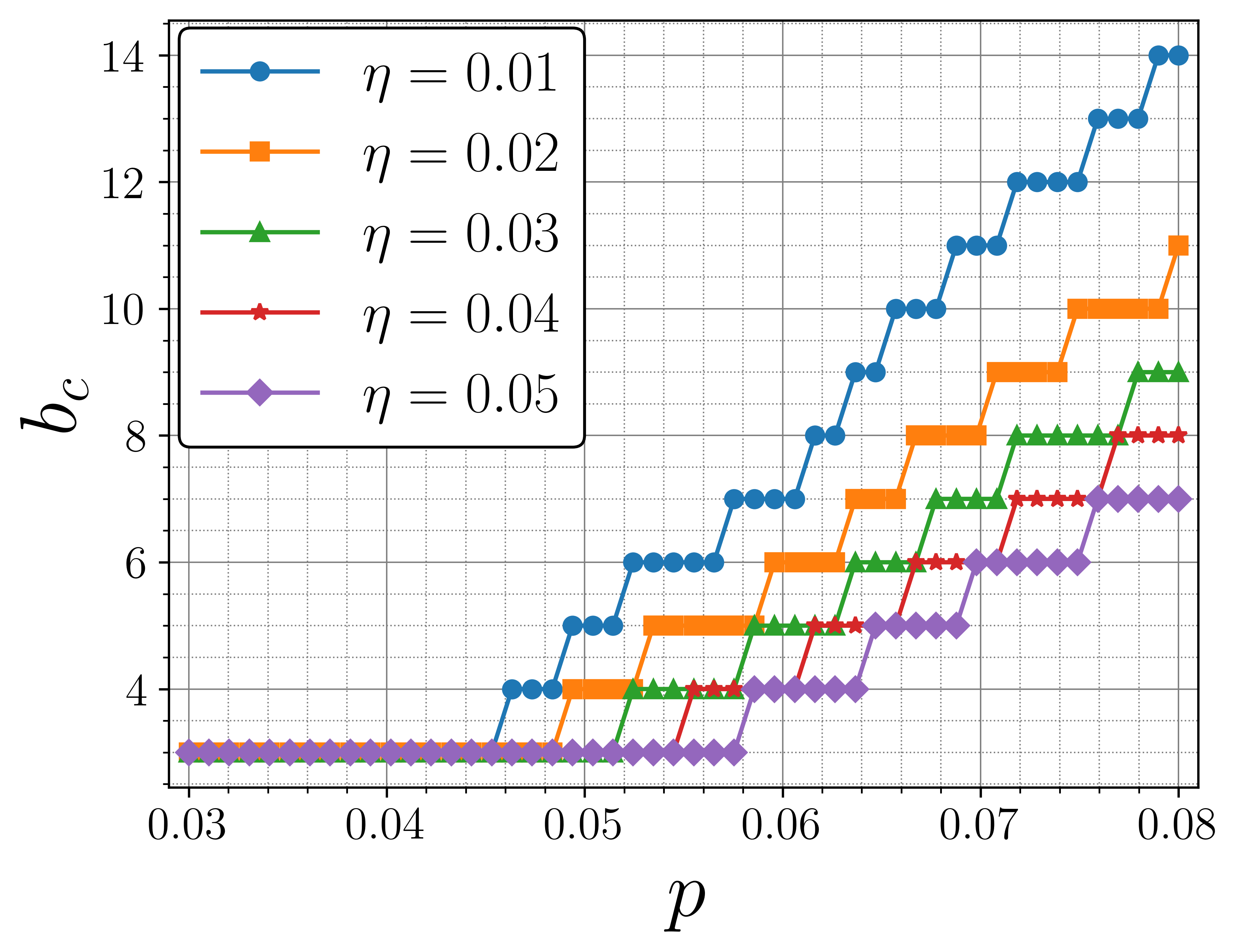}
    \end{minipage}
    \caption{Error-adaptive choice of the buffer size. (a) $R_{\mathrm{match}}$ versus $b$ for representative physical error rates $p$; the dashed line marks the agreement level $R_{\mathrm{match}}=1-\eta=0.98$ corresponding to $\eta=0.02$. (b) Adaptive buffer size $b_c(p,\eta)$ extracted from the same fixed-geometry data, shown for target mismatch tolerances $\eta=0.01$--$0.05$; the required buffer grows as $p$ approaches the transition and as the tolerance $\eta$ is tightened. This analysis motivates the distance-scaled initialization used for the composite decoder below.}
    \label{fig:adaptive_buffer}
\end{figure*}

Far below threshold, a small buffer suffices; as $p$ approaches the transition, the required buffer grows. The matching-ratio scaling thus determines not only that the decoder should be quasi-local, but how quasi-local it must be. Figure~\ref{fig:adaptive_buffer}(a) shows the raw improvement of local-global agreement with increasing buffer size, while panel (b) repackages that information into the operational quantity $b_c(p,\eta;\rho)$ that provides a general initialization rule for the composite decoder.

\section{Composite quasi-local RG decoder}

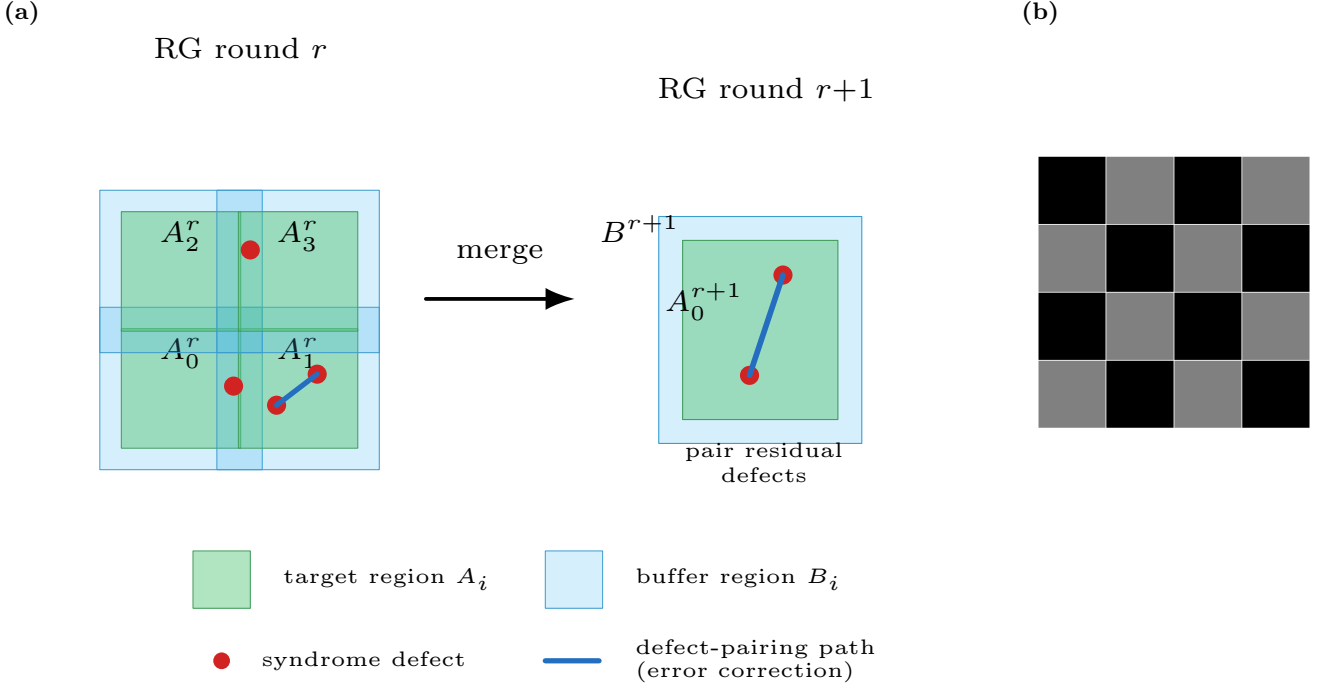
\begin{figure*}[t]
    \centering
    \begin{minipage}[t]{0.74\textwidth}
        \centering
        \makebox[\linewidth][l]{\hspace{0.2em}\textbf{(a)}}\\[-0.10em]
        \resizebox{0.80\linewidth}{!}{\begin{tikzpicture}[x=1cm,y=1cm,>=Latex,line cap=round,line join=round,font=\scriptsize]
    \definecolor{targetfill}{RGB}{82,196,115}
    \definecolor{bufferfill}{RGB}{72,184,242}
    \definecolor{defectred}{RGB}{210,35,35}
    \definecolor{fusionblue}{RGB}{35,110,190}

    \newcommand{\patch}[3]{%
        \begin{scope}[shift={(#1,#2)}]
            \fill[bufferfill,opacity=0.22] (-0.68,-0.68) rectangle (0.68,0.68);
            \draw[bufferfill!85!black,line width=0.24pt] (-0.68,-0.68) rectangle (0.68,0.68);
            \fill[targetfill,opacity=0.45] (-0.50,-0.50) rectangle (0.50,0.50);
            \draw[targetfill!80!black,line width=0.24pt] (-0.50,-0.50) rectangle (0.50,0.50);
            \node at (0,0.30) {$A_{#3}^{r}$};
        \end{scope}
    }

    \node[font=\scriptsize,align=center] at (0.50,2.85) {RG round $r$};
    \patch{0.00}{0.00}{0}
    \patch{0.98}{0.00}{1}
    \patch{0.00}{0.98}{2}
    \patch{0.98}{0.98}{3}

    \fill[defectred] (0.44,0.02) circle (0.080);   
    \fill[defectred] (0.58,1.16) circle (0.080);   
    \fill[defectred] (0.80,-0.14) circle (0.080);  
    \fill[defectred] (1.14,0.12) circle (0.080);   
    \draw[fusionblue,line width=1.25pt] (0.80,-0.14)--(1.14,0.12);

    \draw[->,line width=0.90pt] (2.05,0.75)--(3.30,0.75);
    \node[align=center] at (2.67,1.12) {merge};

    \node[font=\scriptsize,align=center] at (4.90,2.50) {RG round $r{+}1$};
    \fill[bufferfill,opacity=0.22] (4.00,-0.46) rectangle (5.70,1.44);
    \draw[bufferfill!85!black,line width=0.26pt] (4.00,-0.46) rectangle (5.70,1.44);
    \fill[targetfill,opacity=0.45] (4.20,-0.26) rectangle (5.50,1.24);
    \draw[targetfill!80!black,line width=0.26pt] (4.20,-0.26) rectangle (5.50,1.24);
    \node at (4.38,0.73) {$A_{0}^{r+1}$};
    \node at (3.84,1.34) {$B^{r+1}$};

    \fill[defectred] (4.76,0.11) circle (0.080);
    \fill[defectred] (5.04,0.95) circle (0.080);
    \draw[fusionblue,line width=1.35pt] (4.76,0.11)--(5.04,0.95);
    \node[font=\tiny,align=center] at (4.88,-0.62) {pair residual\\defects};

    \newcommand{\legendfont}{\fontsize{4.9}{5.4}\selectfont}
    \fill[targetfill,opacity=0.45] (0.10,-1.84) rectangle (0.58,-1.36);
    \draw[targetfill!80!black,line width=0.24pt] (0.10,-1.84) rectangle (0.58,-1.36);
    \node[anchor=west,font=\legendfont] at (0.74,-1.60) {target region $A_i$};

    \fill[defectred] (0.34,-2.28) circle (0.070);
    \node[anchor=west,font=\legendfont] at (0.56,-2.28) {syndrome defect};

    \fill[bufferfill,opacity=0.22] (3.05,-1.84) rectangle (3.53,-1.36);
    \draw[bufferfill!85!black,line width=0.24pt] (3.05,-1.84) rectangle (3.53,-1.36);
    \node[anchor=west,font=\legendfont] at (3.69,-1.60) {buffer region $B_i$};

    \draw[fusionblue,line width=1.25pt] (3.05,-2.28)--(3.50,-2.28);
    \node[anchor=west,align=left,font=\legendfont] at (3.69,-2.28) {defect-pairing path\\(error correction)};
\end{tikzpicture}}
    \end{minipage}\hspace{0.005\textwidth}
    \begin{minipage}[t]{0.24\textwidth}
        \centering
        \makebox[\linewidth][l]{\hspace{0.2em}\textbf{(b)}}\\[1.20cm]
        \includegraphics[width=\linewidth,trim=28 4 4 4,clip]{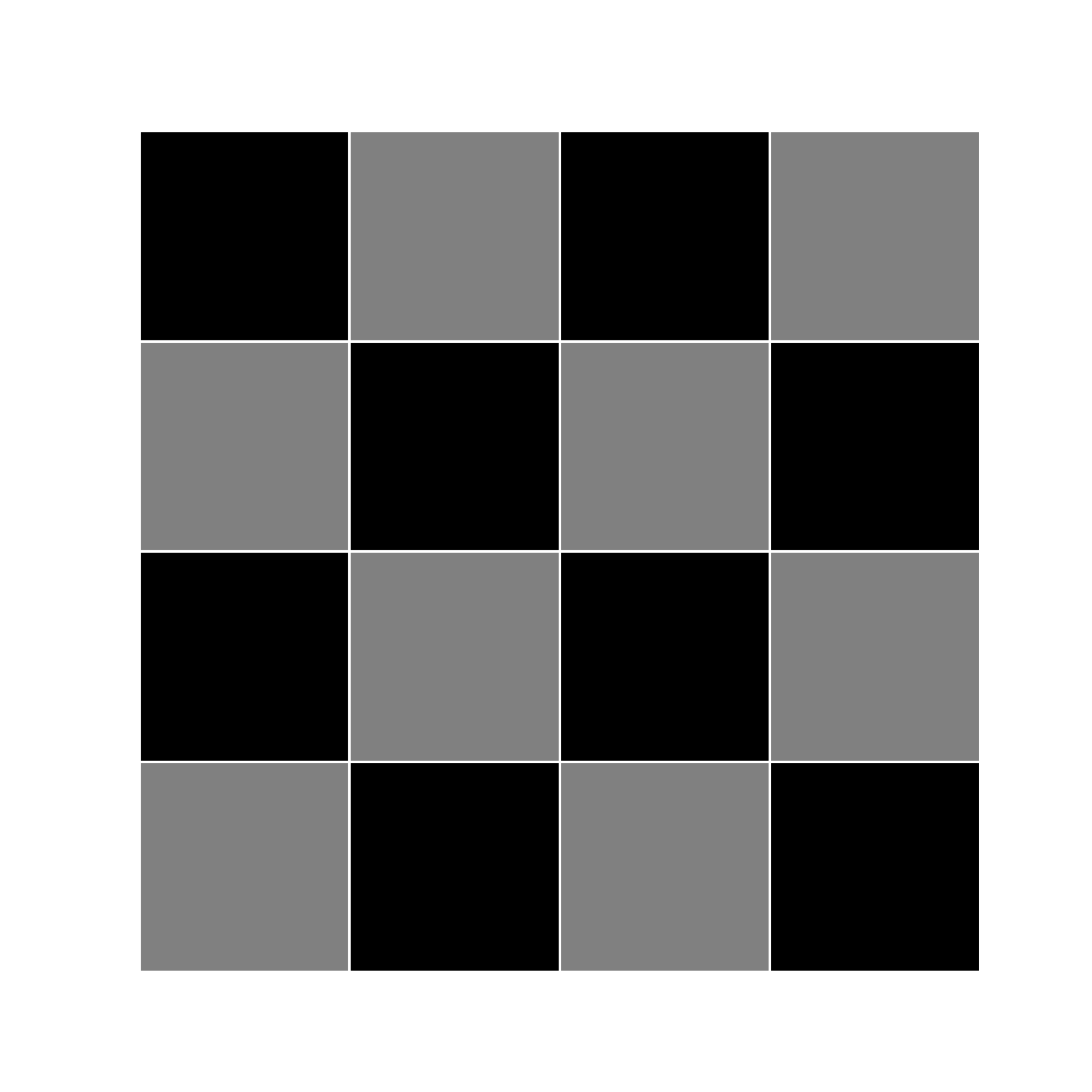}
    \end{minipage}
    \caption{Composite quasi-local decoder architecture. (a) Adjacent local regions $A_i$ with overlapping buffers $B_i$ at RG round $r$, and their merge to a larger region at RG round $(r{+}1)$. Local sectors with no residual defects or with two bulk defects are resolved within RG round $r$, while boundary-associated residual defects are paired after the merge step. Blue segments mark the defect-pairing paths used to annihilate syndrome defects and correct errors. (b) Two-step checkerboard schedule used to parallelize patch updates without double counting on neighboring regions.}
    \label{fig:composite_schematic}
\end{figure*}

With the buffer rule in hand, we now use the local decoder as a building block for a decoder on the whole torus. We refer to this construction as the composite quasi-local RG decoder (composite decoder below). The algorithm is simple in spirit, and one RG round is summarized in Fig.~\ref{fig:composite_schematic}.
\begin{enumerate}
    \item Tile the graph with quasi-local decoders, each defined by a target region $A$ and a surrounding buffer $B$.
    \item Decode each region and keep only the part of the correction acting inside $A$.
    \item Update the syndrome. If residual defects remain, enlarge the local regions and repeat.
    \item Continue this RG-style procedure until no defects remain.
\end{enumerate}

To make the RG part explicit, let $a_r$ and $b_r$ denote the target-region size and buffer width at RG step $r$. The initial choice $(a_0,b_0)$ is taken from the matching-ratio analysis. In particular, one can choose $b_0$ from the target mismatch tolerance through $b_c(p,\eta;\rho)$ at fixed $\rho=a/b$, then set $a_0=\rho b_0$. After each round we enlarge the regions, and the baseline implementation studied here uses a doubling rule,
\begin{equation}
a_{r+1}=2a_r,
\qquad
b_{r+1}=2b_r,
\label{eq:rg_doubling}
\end{equation}
so that neighboring regions with residual defects are merged at the next scale.

One subtlety is that a local patch may contain an odd number of defects, in which case local annihilation inside the patch is not possible. In our implementation this is handled by adding virtual boundary nodes on $\partial(A\cup B)$ in the local MWPM graph, so odd local parity can be absorbed at the boundary and the patch can still return a partial correction. These leftovers, together with residual defects created where correction paths are truncated at the target boundary, are then handled at the next RG step after the regions are enlarged; the RG stage is thus not optional bookkeeping but the part of the decoder that takes care of pairings invisible at the original local scale.

The round-to-round update of $A$ and $B$ is not arbitrary: it is taken from the matching-ratio analysis above, which is how the subregion study informs the whole-system decoder. The local decoder is used to remove the easy short-range part of the syndrome, and the RG enlargement is what promotes the leftover problem to a new local problem on a larger scale.

This architecture is related in spirit to earlier RG decoders \cite{duclos2010fast,bravyi2013quantum,duclos2013fault,duivenvoorden2018renormalization} and, at the level of a single round, to the truncated-MWPM channel of Ref.~\cite{sang2024mixed}, but the organizing principle is different. In this work, each RG round is built from explicit local MWPM patch solves, and the patch geometry is selected from matching-ratio scaling rather than from a fixed coarse-graining prescription. The subregion statistic thus guides the decoder schedule: first remove short-range structure with tuned quasi-local patches, then promote only the residual syndrome to larger scales. That data-driven patch-selection step is the key new ingredient relative to prior RG constructions.

One RG round consists of three steps: decode every active patch, keep only the correction inside its target region, and then recompute the residual syndrome. If the residual syndrome is empty, the algorithm stops. If not, the active regions are merged according to Eq.~(\ref{eq:rg_doubling}) and the same procedure is repeated. The decoder thus interpolates between a strictly local one-step decoder and a global decoder: the first round is local, while later rounds bring in larger and larger scales only where they are needed. Figure~\ref{fig:composite_schematic}(a) illustrates one round of patchwise decoding, the residual-defect patterns that are already resolved locally, and the leftover boundary-associated defects that force the merge to the next scale.

There are also two natural ways to run the decoder. One is sequential, where patches are decoded one after another and the syndrome is updated after each local step. The other is a two-step parallel version based on a checkerboard tiling: first decode all gray regions, then update the syndrome, then decode all black regions. Neighboring buffers can overlap, but only corrections inside each target region $A$ are committed at a given substep, which prevents double counting on shared boundaries. A completely simultaneous update would otherwise over-apply corrections on adjacent regions, so the checkerboard schedule is the minimal way to keep the RG construction parallel without changing the correction rule itself. This gray/black scheduling is closely analogous to the layered structure used in parallel-window decoding, where one first decodes a set of mutually non-overlapping windows in parallel and then uses a second layer of windows to reconcile the regions between them \cite{skoric2023parallel}. In our setting the windows are spatial patches rather than spacetime windows, but the role of the two-coloring is the same.

The checkerboard version is the one relevant for fast decoding. Even though the present implementation is still quasi-local, it already has the right architecture for a parallel decoder, since different patches can be processed simultaneously at each RG stage. The overall cost then depends on three ingredients: the initial quasi-local scale extracted from the matching-ratio ansatz, the cost of the inner patch solver, and the number of RG rounds required before the residual syndrome disappears. This is why the composite decoder should be viewed as an RG architecture rather than as just a local patch decoder run many times.

\section{Results for the composite decoder}

\begin{figure*}[t]
    \centering
    \begin{minipage}[t]{0.47\textwidth}
        \centering
        \includegraphics[width=0.95\linewidth]{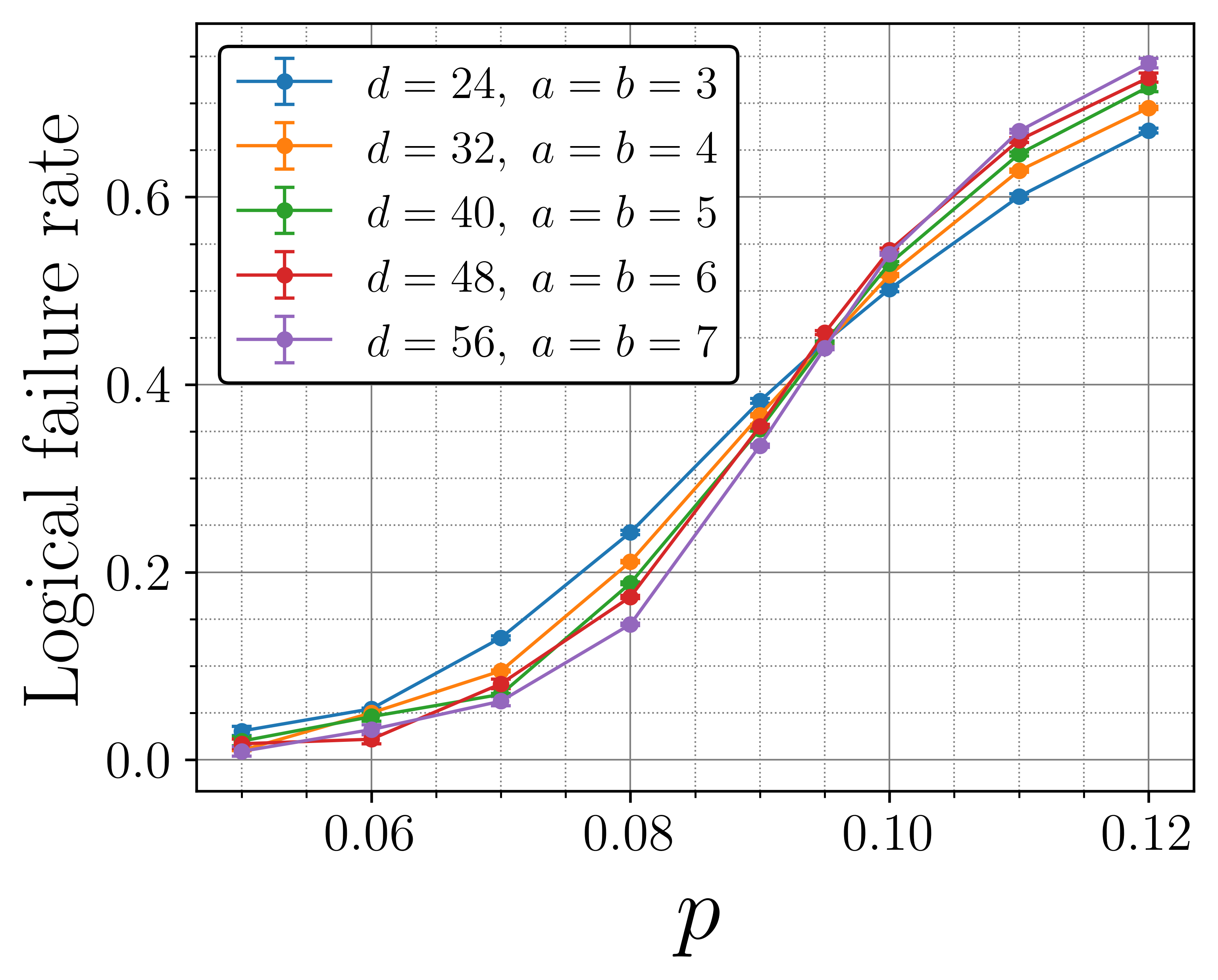}
    \end{minipage}\hspace{0.02\textwidth}
    \begin{minipage}[t]{0.47\textwidth}
        \centering
        \includegraphics[width=0.95\linewidth]{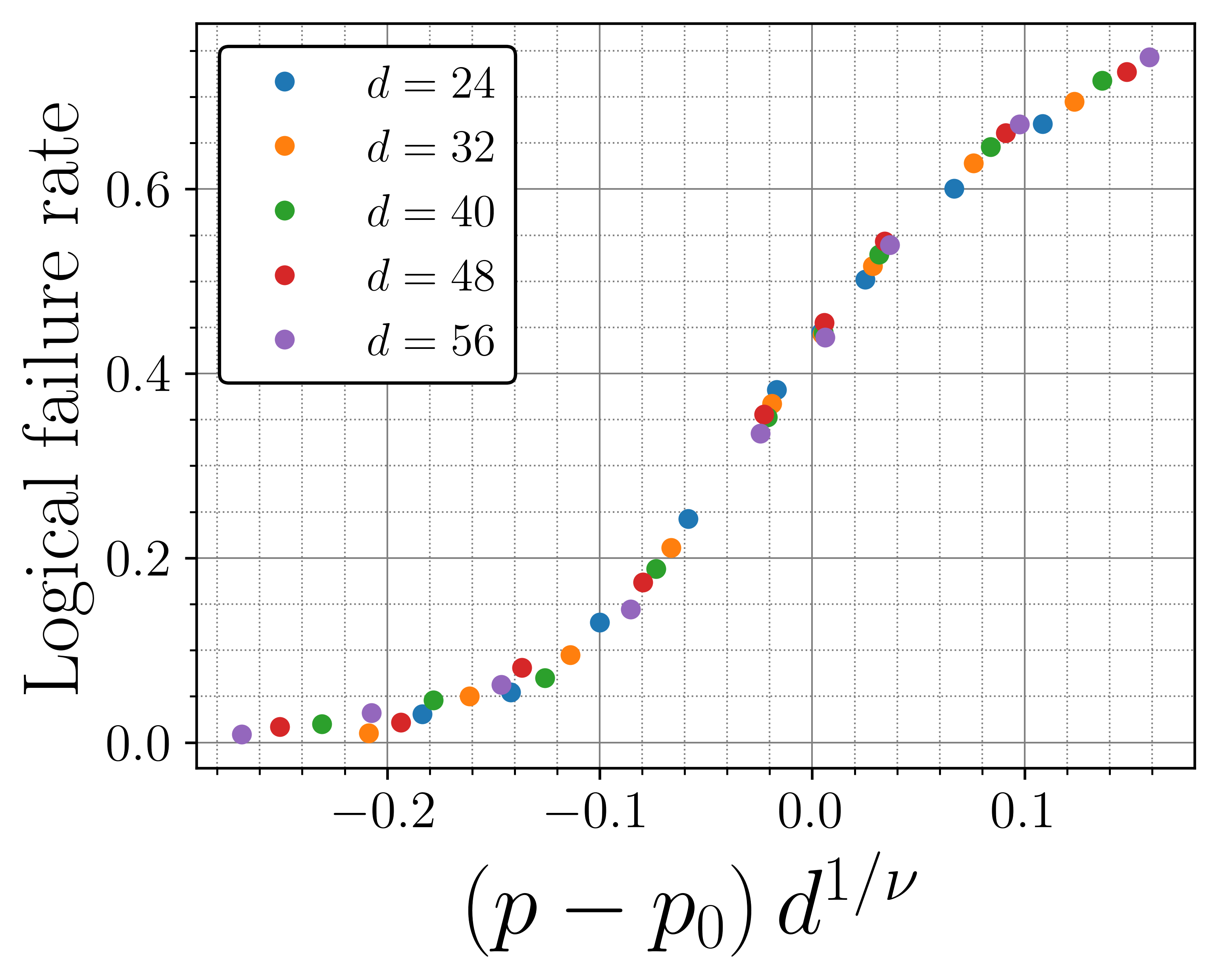}
    \end{minipage}
    \caption{Logical-failure data for the composite quasi-local decoder. For each syndrome sample, the decoder runs local MWPM on active patches, commits only corrections inside each target region $A$, updates the residual syndrome, and, if residual defects remain, merges to the next RG round according to Eq.~(\ref{eq:rg_doubling}) and repeats. The dataset shown here uses the tuned initialization $a_0=b_0=d/8$. Panel (a) shows the resulting logical-failure curves versus $p$, with an apparent crossing window $p_\times^{\mathrm{comp}}\approx 0.09$--$0.10$. Panel (b) shows the finite-size collapse of the same data.}
    \label{fig:whole_system_results}
\end{figure*}

Having constructed the composite decoder from the matching-ratio analysis, we now evaluate its whole-system logical-failure rate. Once the local scales are chosen from the matching-ratio analysis, the whole-system decoder shows threshold-like behavior \cite{watson2014logical}. In the tuned family, we initialize with $a_0=b_0=d/8$ and then apply the RG doubling schedule in Eq.~(\ref{eq:rg_doubling}) until the residual syndrome is eliminated. This distance-scaled initialization is the family analyzed in Sec.~IV; we adopt it uniformly across $p$ rather than re-evaluating $b_c(p,\eta;\rho)$ at each error rate, so the decoder demonstrated here is a fixed scaled-family instance of the general adaptive construction. A direct benchmark of the fully adaptive $b_c(p,\eta;\rho)$ prescription is left for future work. For this setting, the crossing window in Fig.~\ref{fig:whole_system_results}(a) indicates a composite crossing window $p_\times^{\mathrm{comp}}\approx 0.09$--$0.10$, to be compared with $p_c^{\mathrm{MWPM}}\approx 0.103$ for the same noise model.

Matching-ratio scaling therefore does not remain a subregion diagnostic only: it motivates how the local regions are initialized and updated in the composite decoder, and with that choice the whole-system decoder exhibits the threshold-like behavior shown here. At the present sizes we quote only a crossing window. A controlled extraction of the composite threshold with quantified uncertainty would require denser sampling near the crossing and larger code distances, which we leave to future work. We further note that the $d/8$ family was identified and evaluated on the same dataset; selecting the family on a subset of sizes or error rates and evaluating on held-out ones is a natural predictive test that we also leave to future work.

\section{Complexity and fast-decoding implications}

We next examine what the composite construction does and does not imply for fast decoding, distinguishing the multiscale structure it motivates from stronger asymptotic claims that the present data do not yet establish.

In the present implementation, the local patch sizes that work best still grow with the code distance, so the decoder is not strictly local. We are not claiming a constant-depth or proven polylogarithmic-time decoder theorem here.

Still, the structure of the algorithm points in that direction. Let $\ell_r$ denote the typical linear size of the local decoder used at RG step $r$, and let $T_{\mathrm{patch}}(\ell_r)$ be the cost of decoding one such patch. In a checkerboard-parallel implementation, the parallel depth is controlled mainly by the number of RG steps and by the cost of the inner patch solver. Schematically,
\begin{equation}
\mathrm{depth}(d)\sim\sum_{r=0}^{r_{\max}}T_{\mathrm{patch}}(\ell_r),
\qquad
r_{\max}=O(\log d).
\label{eq:depth_sum}
\end{equation}
For blossom-based MWPM on a patch graph with $n_{\mathrm{patch}}=O(\ell_r^2)$ vertices, a conservative worst-case model is $T_{\mathrm{patch}}(\ell_r)=O(n_{\mathrm{patch}}^3)=O(\ell_r^6)$. In practice, modern sparse implementations can be much faster: the sparse-blossom algorithm in PyMatching v2 (the open-source MWPM package used widely in QEC) reports runtime that is approximately linear in detector-graph size over relevant benchmark regimes, while remaining an exact MWPM solver \cite{higgott2023sparse,pymatchingv2}.

If in a future implementation one can keep $\ell_r$ bounded, or at least polylogarithmic in $d$, then the same architecture would lead naturally to
\begin{equation}
\mathrm{depth}(d)=O\!\left(T_{\mathrm{patch}}(\ell_{\max})\log d\right).
\end{equation}
This is the sense in which the present decoder may be relevant for fast decoding: not because the present implementation is already fast in a rigorous asymptotic sense, but because the outer RG structure and the checkerboard schedule are compatible with that goal.

From an applications viewpoint, recent real-time local-decoding programs show that locality can be engineered for high throughput and low latency on dedicated hardware \cite{barber2025realtime,ziad2024local,caune2024realtime,battistel2023real,bombin2023modular,chan2024snowflake}. Combined with an RG-style outer schedule, these ingredients make multiscale implementations of the present architecture technically plausible. Recent studies have also developed local-recoverability and strictly local-decoding viewpoints relevant to this regime \cite{sang2024mixed,sang2024stability,sang2025reversibility,lake2025offline,lake2025active}. Here we use an operational matching statistic to set the quasi-local MWPM patch scaling, and then build the full-system composite decoder from that scaling rule.

\section{Discussion and outlook}

This work turns matching-ratio scaling into a concrete design rule for composite quasi-local RG decoding. The local-global mismatch $\epsilon_{\mathrm{match}}$ follows a simple dependence on $a$ and $b$ that motivates how $A$ and $B$ are initialized and updated across RG rounds. Using this rule, the resulting composite decoder exhibits threshold-like behavior on the full torus. In this framework, subregion analysis guides the decoder geometry and therefore the decoding architecture. The matching ratio serves as an operational recoverability scale, consistent with recent local-recoverability analyses \cite{sang2024mixed,sang2024stability,sang2025reversibility}.

The present implementation is quasi-local and isolates an intermediate regime between fully global and strictly local decoding. A direct next step is to keep the same composite RG structure while replacing the MWPM patch solver with faster local kernels \cite{barber2025realtime,ziad2024local,caune2024realtime,chan2023actis}. The most immediate extension is to noisy syndrome measurements, where decoding becomes intrinsically spacetime and effectively $(2+1)$-dimensional for surface-code memories. In that regime, the local regions are no longer only spatial patches; they become finite spacetime windows, and the matching-ratio diagnostic should be generalized to compare local and global decisions on spacetime worldlines. This extension is also closely related to spacetime-window schemes that parallelize decoding in time by dividing the syndrome history into overlapping windows \cite{tan2023scalable}. Recent information-theoretic and stat-mech analyses of repeated-syndrome dynamics support this direction and suggest concrete diagnostics for finite recoverability scales in the presence of measurement noise \cite{hauser2026information,sang2024stability,sang2025reversibility}.

The same framework can be extended across code families. The most direct step is the surface code with open boundaries \cite{bravyi1998surface}, where edge and corner effects modify the local pairing structure relative to the torus. More broadly, the same local-global agreement statistics can set decoder window and region scales wherever matching-based decoding applies, including higher-dimensional homological constructions \cite{dennis2002topological,bombin2007homological}, color codes decoded by projection to surface-code problems \cite{bombin2006topological,delfosse2014colorprojection,kubica2015unfolding}, and bivariate-bicycle codes with matching-based decoders \cite{mackay2004sparsegraph,bravyi2024highthreshold,sahay2026matchingbb,tan2026generalizedmatching}. A complementary longer-term direction is to promote the elementary objects of the decoder from physical patches to logical-gadget spacetime blocks, connecting to recent correlated-decoding and transversal-logic analyses \cite{cain2024correlated,zhou2025lowoverhead,cain2025fast,xu2026rigorous}.

\begin{acknowledgments}
H.D. was supported in part by NSF grant OMA-2120757, and Simons Foundation. M.G. was supported by the NSF funded NQVL:QSTD: Design: QRAQL, and NSF grant PHY-2309135 to the Kavli Institute for Theoretical Physics (KITP). The authors acknowledge the University of Maryland supercomputing resources (https://hpcc.umd.edu) made available for conducting the research reported in this paper.
\end{acknowledgments}

\appendix

\section{Simulation methods and decoder algorithms}

For completeness, this appendix records the algorithmic details that underlie the numerical implementation used in the main text: the patch-level update rule and the full composite RG decoder, in compact form.

The simulations use toric-code syndrome snapshots under dephasing noise with perfect syndrome measurements. Let $s\in\mathbb{F}_2^{m}$ be the syndrome vector, $c\in\mathbb{F}_2^{n}$ the accumulated Pauli-$Z$ correction vector on edges, and $H$ the parity-check matrix for $X$ stabilizers. Syndrome updates are performed as
\begin{equation}
s \leftarrow s \oplus H\Delta c
\end{equation}
after each committed correction increment $\Delta c$.

Patch geometry is defined by a target square $A$ (linear size $a_r$) and buffer width $b_r$ at RG round $r$. The initial $(a_0,b_0)$ is selected from the matching-ratio analysis in the main text; subsequent rounds follow Eq.~(\ref{eq:rg_doubling}). The parallel implementation uses a checkerboard schedule so that no two simultaneously updated target regions share a boundary.

\onecolumngrid
\begin{algorithm}[H]
\caption{\textsc{PatchMWPM}: local MWPM patch update}
\label{alg:local_patch_update}
\begin{algorithmic}[1]
\Require Current syndrome $s$, target region $A$, buffer $B$, local graph on $A\cup B$, deterministic MWPM tie-breaking
\Ensure Committed local correction increment $\Delta c_A$ (supported on $A$)
\State Restrict syndrome defects to $A\cup B$ and build the local matching graph.
\State Add virtual boundary nodes on $\partial(A\cup B)$ so odd local defect parity is allowed.
\State Run MWPM on the local graph to obtain $\Delta c_{A\cup B}$.
\State Project to target support: $\Delta c_A \gets \Pi_A(\Delta c_{A\cup B})$.
\State \Return $\Delta c_A$
\end{algorithmic}
\end{algorithm}

\begin{algorithm}[H]
\caption{Composite quasi-local RG decoder}
\label{alg:composite_rg_decoder}
\begin{algorithmic}[1]
\Require Initial syndrome $s_0$, maximum rounds $r_{\max}$, initial geometry $(a_0,b_0)$, mode $\in\{$sequential, checkerboard$\}$
\Ensure Final correction $c$ and residual syndrome $s$
\State $c \gets 0$, $s \gets s_0$, $r \gets 0$
\State Build initial active patch set $\mathcal{P}_0$ from $(a_0,b_0)$
\While{$s \neq 0$ and $r \le r_{\max}$}
    \If{mode is checkerboard}
        \State Build ordered groups $\mathcal{G}_r \gets \{$gray patches in $\mathcal{P}_r$, black patches in $\mathcal{P}_r\}$
    \Else
        \State Build ordered groups $\mathcal{G}_r \gets \{\{P\}: P \in \mathcal{P}_r\}$ (one patch per group, in update order)
    \EndIf
    \ForAll{groups $g \in \mathcal{G}_r$ in update order}
        \ForAll{patches $P \in g$}
            \State $\Delta c_{A(P)} \gets$ \Call{PatchMWPM}{$s,A(P),B(P)$}
        \EndFor
        \State $\Delta c_g \gets \bigoplus\limits_{P \in g}\Delta c_{A(P)}$
        \State $c \gets c \oplus \Delta c_g$
        \State $s \gets s \oplus H\Delta c_g$
    \EndFor
    \If{$s = 0$}
        \State \Return $(c, s)$
    \EndIf
    \State Enlarge/merge active regions via Eq.~(\ref{eq:rg_doubling}) to form $\mathcal{P}_{r+1}$
    \State $r \gets r+1$
\EndWhile
\State \Return $(c, s)$ \Comment{$r_{\max}$ safety cap reached; unresolved syndrome indicated by $s\neq 0$}
\end{algorithmic}
\end{algorithm}
\twocolumngrid

In practice we use deterministic solver settings so repeated runs on identical syndromes produce identical patch-level outputs, which is necessary for stable matching-ratio statistics and reproducible checkerboard updates. In our simulations, $r_{\max}$ is used as a safety cap, not as a design ingredient of the decoder schedule; within the parameter ranges reported here, this cap was not saturated.

\begin{figure}[t]
    \centering
    \includegraphics[width=0.456\linewidth]{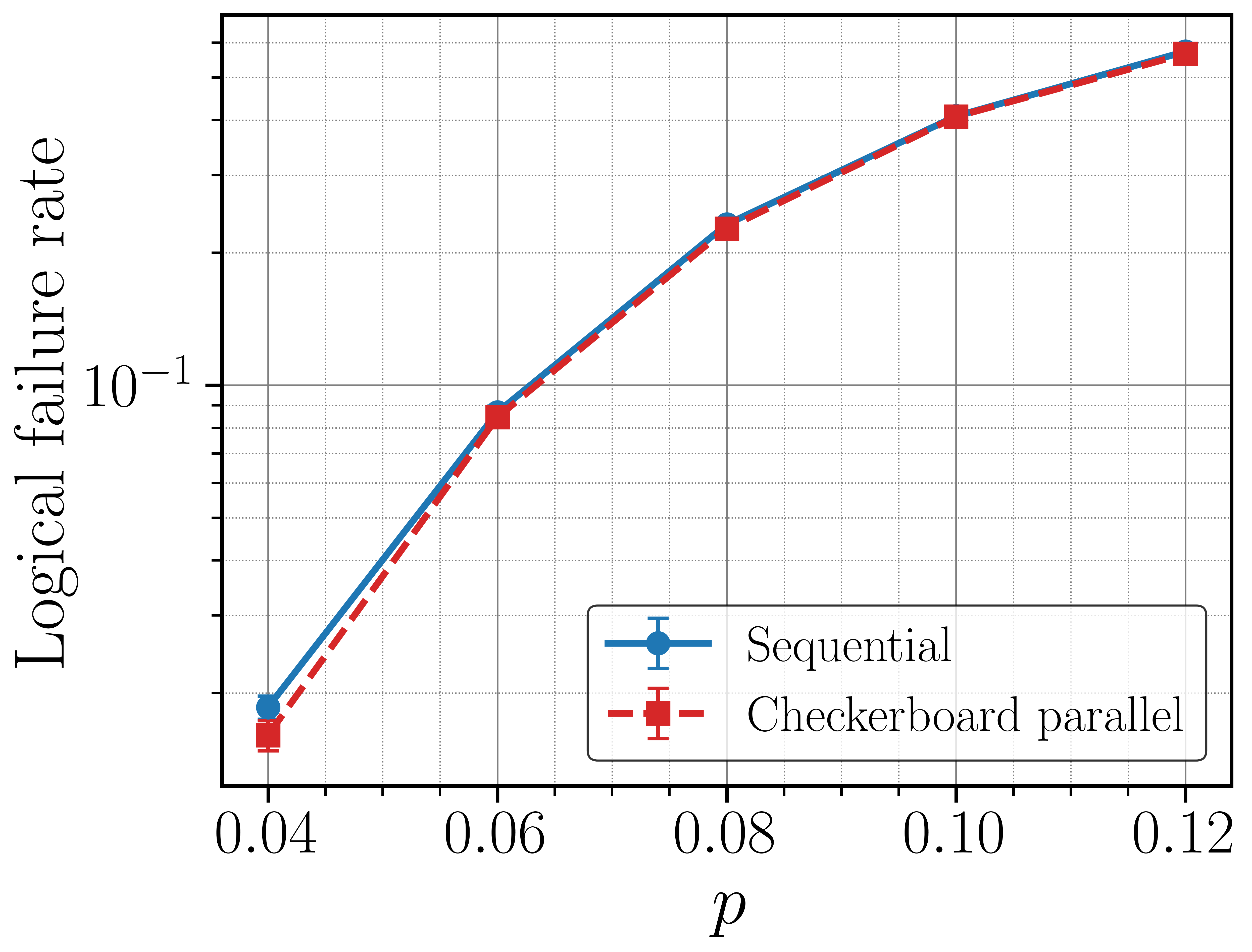}
    \caption{Appendix control: checkerboard-parallel versus sequential composite-decoder updates for matched geometry $(d,a,b)=(12,4,4)$ and matched shot budget. The horizontal axis is the physical error rate $p$. The two update schedules track each other across the analysis window.}
    \label{fig:appendix_series_parallel}
\end{figure}

\begin{figure}[t]
    \centering
    \begin{minipage}[t]{0.49\linewidth}
        \centering
        \includegraphics[width=0.95\linewidth]{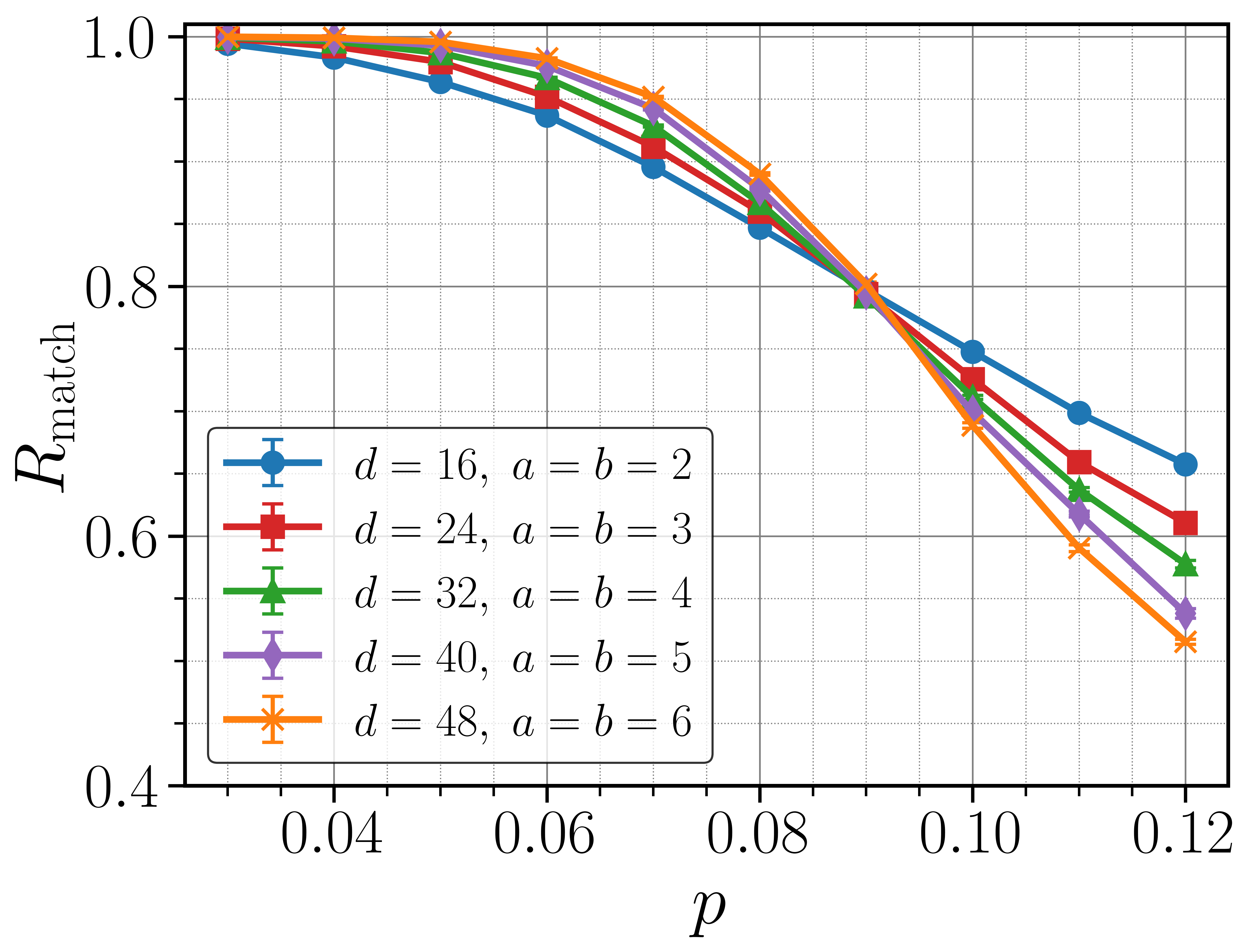}
    \end{minipage}\hfill
    \begin{minipage}[t]{0.49\linewidth}
        \centering
        \includegraphics[width=0.95\linewidth]{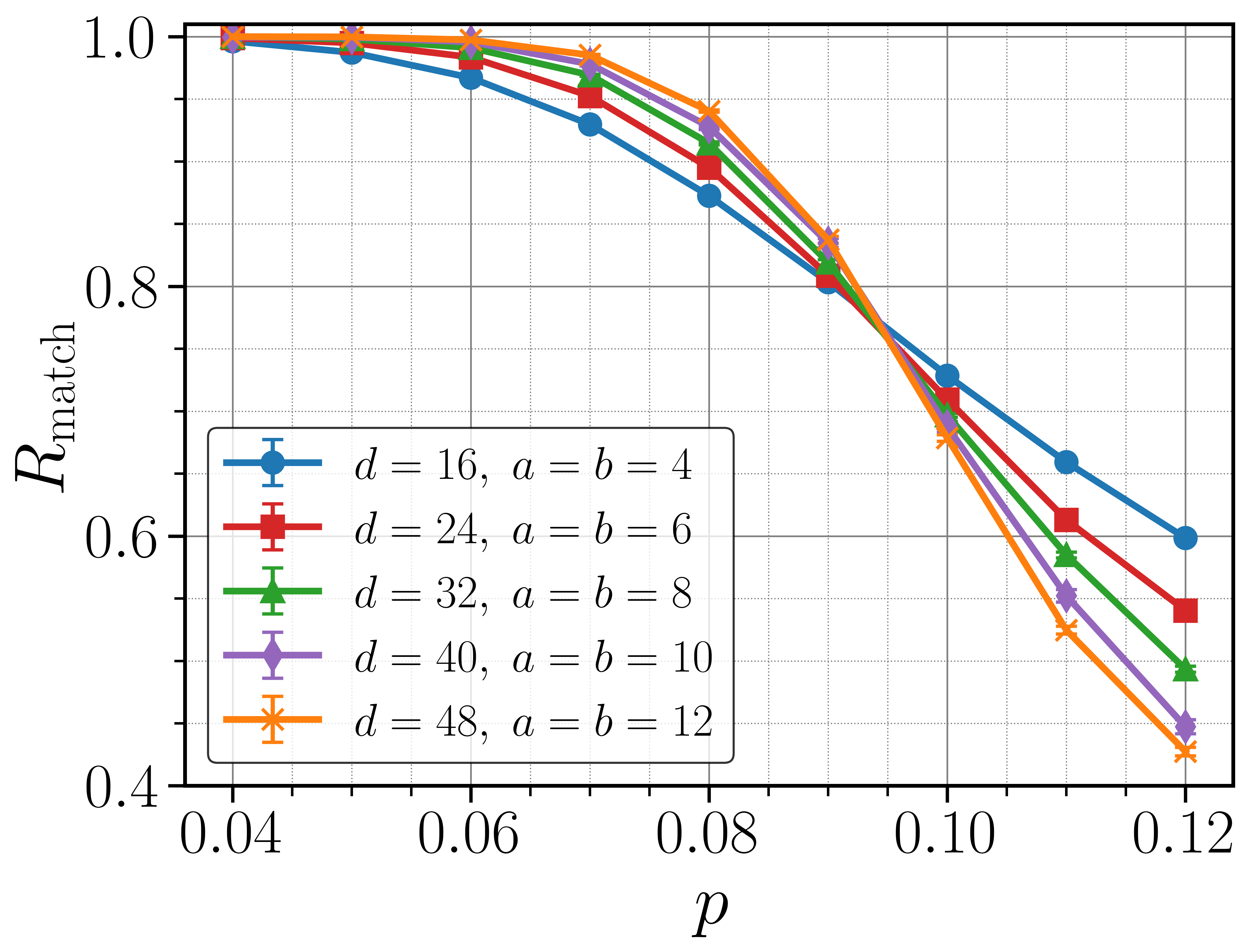}
    \end{minipage}
    \caption{Appendix control: additional distance-scaled matching-ratio families at aspect ratios $a=b=d/8$ (left) and $a=b=d/4$ (right), quoted in the same vertex-lattice units used throughout. The horizontal axis is the physical error rate $p$. The left panel reproduces the main-text $a=b=d/8$ family on an independent set of code distances, and the right panel exhibits the same qualitative crossing for the larger $a=b=d/4$ family.}
    \label{fig:appendix_scaled_families}
\end{figure}

We also include two supplementary controls with matched shot budgets. Figure~\ref{fig:appendix_series_parallel} compares checkerboard-parallel and sequential update schedules at a single matched geometry, and Fig.~\ref{fig:appendix_scaled_families} shows the same crossing trend at aspect ratios $a=b=d/8$ (on an independent set of code distances) and $a=b=d/4$ (a larger target-to-buffer scaling).

\onecolumngrid

\twocolumngrid

\clearpage
\bibliographystyle{apsrev4-2}
\bibliography{Ref}

\end{document}